\documentclass[11pt]{article}

\usepackage{color}
\usepackage{graphicx}
\usepackage{subfig}
\usepackage{booktabs}
\usepackage{tikz}
\usetikzlibrary{decorations.pathmorphing}
\usetikzlibrary{decorations.pathreplacing}
\usepackage{amsmath}
\usepackage{url}

\DeclareMathOperator*{\argmax}{arg\,max}
\newcommand{\nrAffStrings}{159,068} 
\newcommand{\nrAffStringsGeo}{117,942} 
\newcommand{\nrAffsCP}{4,254} 
\newcommand{\nrPubs}{1,894,758} 
\newcommand{\nrAuthors}{1,080,958} 
\newcommand{\nrMobileAuthors}{193,986} 
\newcommand{\nrAuthorsReturned}{29,398} 
\newcommand{\nrInitLabeledAuthAffs}{479,258} 
\newcommand{\nrLabeledAuthAffs}{4,318,206} 

\newcommand{\nrOrganizationsMS}{13,276} 
\newcommand{\nrHops}{310,282}

\begin{document}

\title{GeoDBLP: Geo-Tagging DBLP for Mining the Sociology of Computer Science}

\author{
Fabian Hadiji$^{1,2}$ \ \ Kristian Kersting$^{1,2}$ \\ Christian Bauckhage$^{1,2}$ \ \ Babak Ahmadi$^{2}$\\
{\small $^1$University of Bonn, Germany \ \ $^2$Fraunhofer IAIS, Germany}\\
\texttt{\small \{firstname.lastname\}@iais.fraunhofer.de}
}
\date{}
\maketitle

\begin{abstract}
Many collective human activities have been shown to exhibit
universal patterns. However, the possibility of universal patterns
across timing events of researcher migration has barely been explored at
global scale. Here, we show that timing events of migration
within different countries exhibit remarkable similarities. Specifically, 
we look at the distribution governing the data of researcher
migration inferred from the web. Compiling the data in itself represents
a significant advance in the field of quantitative analysis of migration
patterns. Official and commercial records are often access restricted,
incompatible between countries, and especially not registered across researchers.
Instead, we introduce GeoDBLP where we propagate geographical seed locations 
retrieved from the web across the DBLP database of 1,080,958 authors and 1,894,758 papers. 
But perhaps more important is that we are able to find
statistical patterns and create models that explain the
migration of researchers. For instance, we show that the science
job market can be treated as a Poisson process with
individual propensities to migrate following a 
log-normal distribution over the researcher's career stage. 
That is, although jobs enter the market constantly, researchers are generally not ``memoryless'' but have to
care greatly about their next move. The propensity to make $k>1$ migrations, however,
follows a gamma distribution suggesting that migration at later
career stages is ``memoryless''. This 
aligns well but actually goes beyond scientometric
models typically postulated based on small case studies. On a very large,
transnational scale, we establish the first general regularities that should have major 
implications on strategies for education and research worldwide.

\end{abstract}

\section{Introduction}
Over the last years, many collective human activities have been shown to exhibit
universal patterns, see e.g.~\cite{zipf46asr,mantegna95nature,gabaix03nature,barabasi05nature,cohen08pnas,hidalgo08physica,leskovec08www,gonzales08nature, bohorquez09nature,leskovec09kdd,goetz09icwsm,wang11kdd,barabasi12nature} among 
others. 
However, the possibility of universal patterns
across timing events of researcher migration --- the event of transfer from one residential location
to another by a researcher --- has barely been explored at
global scale. This is surprising since education and science is, and has always been international. 
For instance, according to the UNESCO Institute for Statistics, the global number of foreign students
pursuing tertiary education abroad increased from $1.6$ million in 1999 to $2.8$ million in
2008\footnote{United Nations Education, Scientific and Cultural Organization, Data extract (Paris, 2011), accessed on 19 April
2011 at: \url{http://stats.uis.unesco.org/unesco/TableViewer}}. As the UN notes~\cite{un2012}, ``there has been an expansion of arrangements whereby universities from
high-income countries either partner with universities in developing countries or establish
branch campuses there. Governments have supported or encouraged these arrangements,
hoping to improve training opportunities for their citizens in the region and to attract
qualified foreign students.'' Likewise, 
science thrives on the free exchange of findings and methods, and ultimately of the researchers themselves, as noted by the 
German Council of Science and Humanities~\cite{wissenschaftsrat10}. The 
European Union even defined the free movement of knowledge in Europe as the ``fifth fundamental freedom''~\footnote{Council of the European Union (2008a), p. 5:
``In order to become a truly modern and competitive
economy, and building on the work carried out on the future of science and technology and on the modernization
of universities, Member States and the EU must remove barriers to the free movement of knowledge
by creating a `fifth freedom' ...''}. Similarly, the US National Science Foundation argues that  
``international high-skill migration is likely to have a positive effect on global incentives for human capital investment. 
It increases the opportunities for highly skilled workers both by providing the option to search for a job across borders 
and by encouraging the growth of new knowledge''~\cite{nsf07}. Generally, due
to globalization and rapidly increasing international competition,
today's scientific, social and ecological challenges
can only be met on a global scale both in education and science,  
and are accompanied by political and economic interests. Thus, research on
scientist's migration and understanding it, play key roles in the 
future development of most computer science departments, research institutes, companies and 
nations, especially if fertility continues to decline globally~\cite{lee11science}. 
But can we provide decision makers and analysts with statistical regularities of migration? 
Are there any statistical patterns at all? 

These questions were the seed that grew into the present report. On first sight, reasons to migrate 
are manifold and complex: political stability 
and freedom of science, family influences such as long distance relationships 
and oversea relatives, and personal preferences such as exploration, climate, improved career, 
better working conditions, among others. Despite this complex web of interactions, 
we show in this paper that {\it the timing events of migration
within different countries exhibit remarkable simple but strong and similar regularities.} 
Specifically, we look at the distribution governing the data of researcher
migration inferred from the web. Compiling the data in itself represents
a significant advance in the field of quantitative analysis of migration
patterns. Although, efforts to produce comparable and reliable statistics are underway,
estimates of researcher flows are inexistent, outdated, or largely inconsistent, 
for most countries. Moreover, official (NSF, EU, DFG, etc.) and commercial (ISI, Springer, Google, AuthorMapper, ArnetMinder) 
records are often access restricted and especially not registered across researchers. 
On top of it, these information sources are often highly noisy.
Luckily, bibliographic sites on the Internet such as DBLP are publicly accessible and contain data for millions of publications. 
Papers are written virtually everywhere in the scientific world, and the affiliations of authors tracked over time could be used as 
proxy for migration. Unfortunately, many if not most of the prominent bibliographic sites 
such as DBLP do not provide affiliation information. 
Instead, we have to infer this information. 
To do so, we extracted the geographical locations --- the cities --- for a few 
seed author-paper-pairs and then propagated them across the DBLP social network 
of more than one million authors and almost two million papers. 
We refer to this new dataset as {\bf GeoDBLP}, DBLP augmented with geo-tags.
GeoDBLP is the basis for our statistical analysis and has city-tags for most of 
the 5,033,018 paper-author-pairs in DBLP. Specifically, as partly shown in 
Fig.~\ref{fig:main}, we present the first strong regularities for researcher migration in computer science: 

\begin{figure}[h]
\includegraphics[width=.22\textwidth, page=3]{figs/sketches_time_to_hop}
\hfill
\includegraphics[width=.22\textwidth, page=2]{figs/sketches_time_to_kth_hop_4}
\hfill
\includegraphics[width=.22\textwidth, page=2]{figs/sketches_time_to_return}
\hfill
\includegraphics[width=0.22\textwidth]{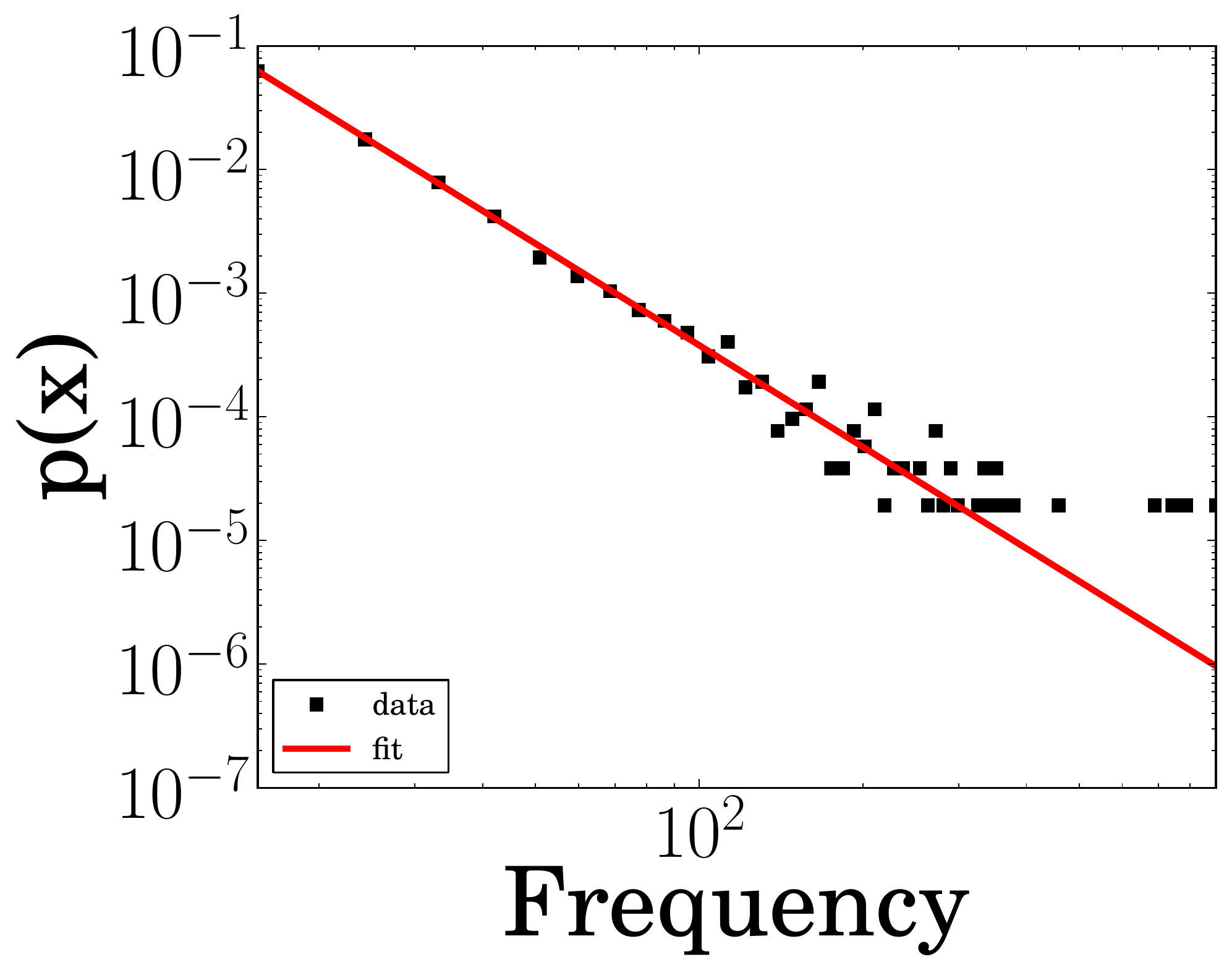}
\caption{We infer from the WWW the first strong regularities of timing events in the migration of computer scientists. Due to the many early stage careers, with non-permanent contracts, a specific scientist's propensity to make the next move follows a log-normal distribution (left). For larger numbers of moves, i.e., for senior scientists this turns into a gamma distribution due to permanent positions (left-middle); migration becomes memoryless. The circulation of expertise, i.e., the time until a researcher returns to the country of her first publication follows a gamma distribution (middle-right). Returning is also memoryless. The inter-city
migration frequency distribution, however, follows a power-law (right). That is, cities with a high exchange of researchers will even exchange more researchers in the future. These regularities should have major implications on
strategies for research across the world.\label{fig:main}}
\end{figure}

\begin{itemize}
\item {\bf (R1)} A specific researcher's propensity to migrate, that means to make the next move, follows a log-normal distribution. 
That is, researchers are generally not ``memoryless'' but have to care 
greatly about their next move. 
This is plausible due to the dominating early career researchers with
non-permanent positions. 
This regularity of timing events is remarkably stable and similar within different continents and countries across the globe.
\item {\bf (R2)} The propensity to make $k>1$ migrations, however,
follows a gamma distribution suggesting that migration at later
career stages is ``memoryless''. 
That is, researchers have to care less about their next move since the majority of positions are permanent in later career stages. 
\item Since jobs enter the market all the time, R1 and R2 together suggest that
the job market can be treated as a Poisson-log-normal process. 
\item {\bf (R3)} The brain circulation, i.e., the time until a researcher returns to the country of her first publication, follows a gamma distribution.
That is, returning is also memoryless. 
Researchers cannot plan to return but rather have to pick up opportunities as they arrive.  
\item {\bf (R4)} The inter-city migration frequency follows a power-law. That is, cities with a high exchange of researchers will exchange even more researchers
in the future. So, investments into migration pay off. 
\item {\bf Statistical patterns:} Link analysis of the author-migration graph can discover additional statistical patterns such as {\bf (SP1)} migration sinks, sources and incubators, as well as {\bf (SP2)} the hottest migration cities. 
\end{itemize}
These results validate and go beyond migration models based on small case 
studies at a very large, transnational scale. Ultimately, they can provide forecasts of 
(re-)migration which can help decision makers who seek actively the migration and the return of 
their researchers to reach better decisions regarding the timing of their efforts.

Already Zipf~\cite{zipf46asr} investigated inter-city migration. He analyzed so called gravity models.
These models incorporated terms measuring the masses of each origin and destination and the distance
between them and were calibrated statistically using log-linear regression techniques.
Over the years, several modifications and alternatives have been postulated, see e.g.~\cite{cohen08pnas,Siminil12nature} and the references in there.  
Steward~\cite{stewart94info} reviewed the Poisson-log-normal model for bibliometric/scientometric
distributions, i.e., to characterize the productivity of scientists. Sums of 
Poisson processes and other Poisson regression models as well as ordinary-least-squares 
have actually a long tradition within migration research, see \cite{stillwell05,rees09} 
for recent overviews. All of these
approaches, however, have considered small scale data only~\cite{rees09} and have not considered researcher
migration in computer science. To the best of our knowledge, the 
only large-scale migration study 
was recently presented by Zagheni and Weber~\cite{zagheni12websci}, analyzing a large-scale  
e-mail datasets to estimate international migration rates, but not specific to computer scientists. Moreover, they have not presented any statistical regularities nor
dealt with missing information. Indeed, as already mentioned, other collective human activities
have been the subject of extensive and large-scale planetary mining. Prominent examples are 
mobility patterns drawn from communication~\cite{leskovec08www,hui10hot} and web services~\cite{noulas11wscm}, 
as well as mining blog dynamics~\cite{goetz09icwsm} and social ties~\cite{wang11kdd}.  
Our methods and findings complement these results by highlighting the value of
using the World Wide Web together with data mining to deal with missing information 
as a world-wide lens onto researcher migration,
enabling the analyst to develop global strategies for
research migration and to inform the public debate. 

We proceed as follows. We start by discussing the harvesting of
our data in detail. Then, we will describe how we made use of multi-label propagation to
fill in missing information. Before concluding, we will present our statistical migration 
models and patterns. 

\section{Mining the Data from the Web}
Bibliographic sites on the Internet such as DBLP are publicly accessible and  contain millions of data records on publications. 
Papers are written virtually everywhere in the scientific world, and the affiliations of authors tracked over time could be used as 
proxy for migration. Unfortunately, many if not most of the prominent bibliographic sites such as DBLP
do not provide affiliation information. Instead we have to infer this information. 
In this section, we will detail the mining of our data. 
The goal was to tag every of the over 5 million
author-paper-pairs in our database with an affiliation. 
The data collection method utilized an open-source information extraction 
methodology, namely  DBLP, ACM Digital Library, Google's Geocoding API and 
large-scale multi-label propagation.


\subsection{Harvesting the Data}

\begin{figure}[t]
\centering
\subfloat[Publications per year]{
\includegraphics[width=.35\columnwidth]{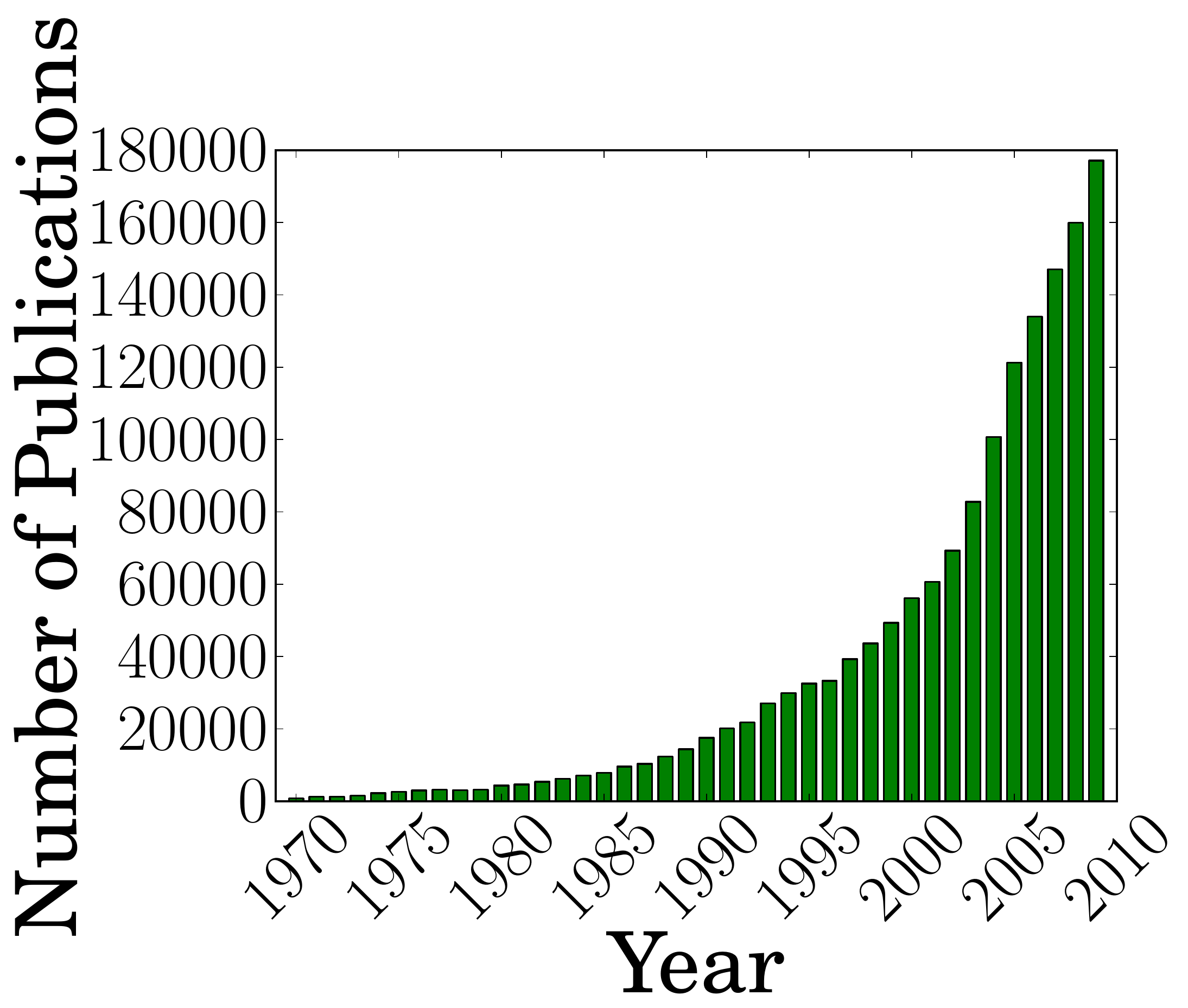}
\label{fig:dblp_pubs}
}
\qquad\qquad
\subfloat[Authors per year]{
\includegraphics[width=.35\columnwidth]{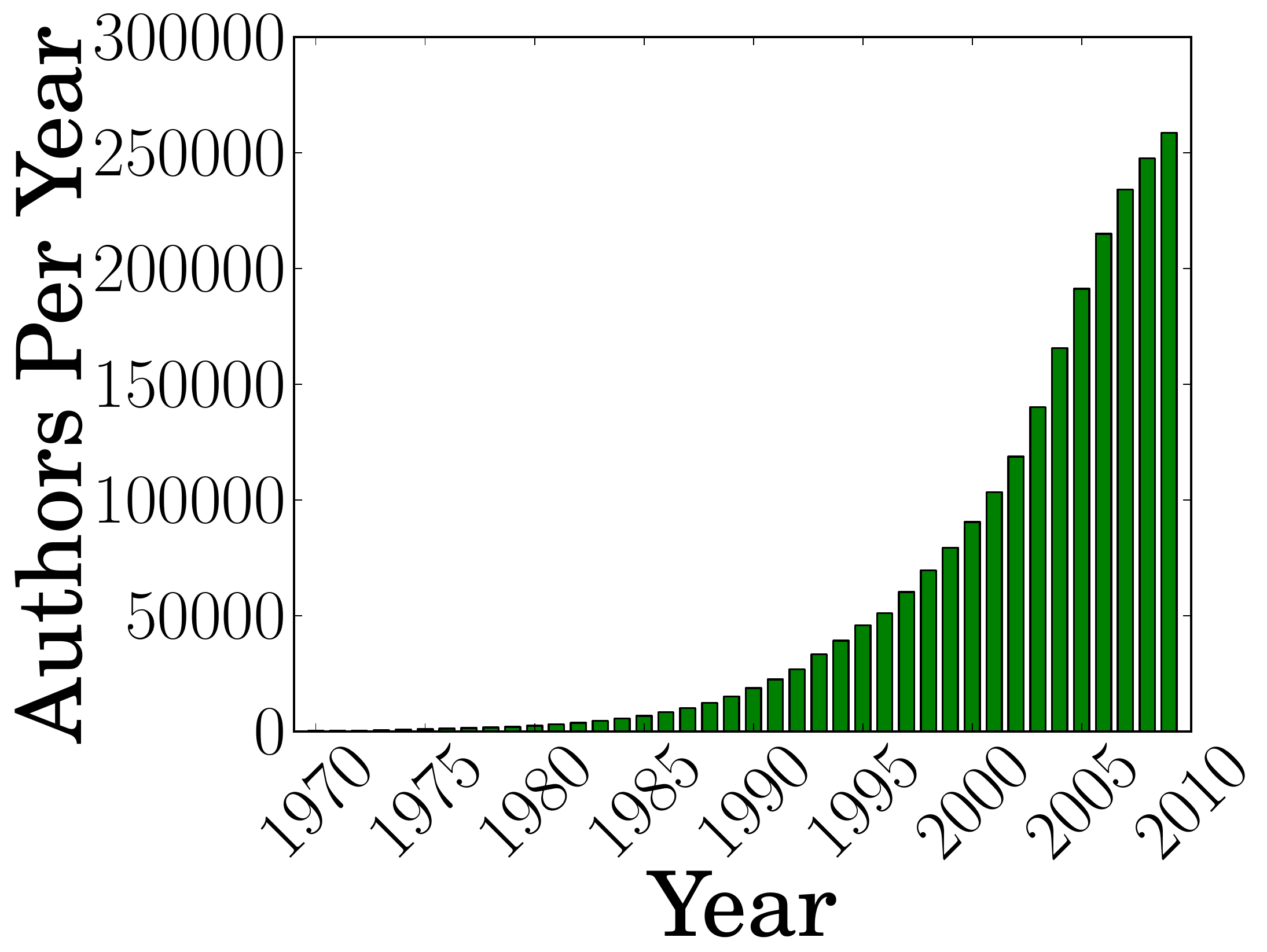}
\label{fig:dblp_authors}
}
\caption{Statistics of the DBLP dump: The number of publications (\ref{fig:dblp_pubs}) and authors (\ref{fig:dblp_authors}) per year.
 As one can see, DBLP has been growing constantly over the past decades from 1970 until 2010. 
\label{fig:dblp_stats}}
\end{figure}

We used DBLP\footnote{\url{http://dblp.uni-trier.de/}} as a starting point. 
DBLP is a large index of computer science publications that also offers a manual best-effort entity disambiguation~\cite{ley09}. 
We used an XML-dump from February 2012 which contained \nrPubs{} publications written by \nrAuthors{} authors.
Fig.~\ref{fig:dblp_stats} shows the number of publications and authors per year from this dump. As one can see,
the number of computer scientists as well as the productivity have been growing enormously over the past decades.
Unfortunately, DBLP does not provide affiliation information for the authors over the years. This information, however,
is required in order to develop migration models using author affiliations as proxy. 
Specifically, we aim to infer geo-tags of the more than 5 million unknown author-paper-pairs.

Luckily, there are other information sources on the web that contain such information.
One of these systems is the ACM Digital Library\footnote{\url{http://dl.acm.org/}}. Unfortunately, 
ACM DL does not allow a full download of the data. 
Consequently, we retrieved the affiliation information of only a few 
papers from ACM DL which we then had to match with our DBLP dump. 
This resulted in affiliation information for \nrInitLabeledAuthAffs{} of all author-paper-pairs. 
In order to fill in the missing information, we resorted to data mining techniques. 
To do so, however, we have to be a little bit more careful. 
First, the names of the affiliations in ACM DL are not in canonical form which results in a very large set of affiliation candidates. 
More precisely, the DBLP dump enhanced with the initial affiliations from ACM DL contained \nrAffStrings{} different affiliation names in total.
Secondly, although we have now partial affiliation information, we still lack exact geo-information of the organizations to identify cities, countries, and continents.
Many of the affiliation names may contain a reference to the city or country but these pieces of information are not trivial to extract from the raw strings.
Additionally, we want to have latitude and longitude values to enable further analysis and visualization.
For example, latitude and longitude data would allow one to calculate exact distances between collaborators.
This geo-location issue can easily be resolved using Google's Geocoding API\footnote{\url{https://developers.google.com/maps/documentation/geocoding/}}.
Just querying the API using the retrieved affiliation names 
resulted in geo-locations for \nrAffStringsGeo{} of the \nrAffStrings{} strings. 
The remaining gap primarily rises from the fact that the Google~API does not 
find geo-locations for all the retrieved affiliation strings. 
This is essentially because the strings contain information not related to the geo-location such as departments, e-mail addresses, among others. 
In any case, as our empirical results will show, this resulted in enough information to propagate the seed affiliations and in turn the geo-locations across the DBLP network of authors and papers. 

\subsection{Inferring Missing Data}

Before we infer the missing author-paper-pairs, we revise our obtained affiliation data. To further increase the quality of our harvested affiliations, we hypothesized that there are actually not that many relevant organizations in Computer Science and these names need to get de-duplicated. 
This hypothesis is confirmed by services such as {\it MS Academic Search}\footnote{\url{http://academic.research.microsoft.com/}} which currently lists only \nrOrganizationsMS{} organizations compared to our 150k+ names.
Since, we now have the geo-locations for many of the affiliation strings, we can use this information for a simple entity resolution which helps resolving this issue.
More precisely, we clustered affiliations together for which the retrieved city coincide resulting in \nrAffsCP{} distinct cities\footnote{Indeed, this approach does not distinguishing multiple affiliations per cities such as {\it MIT} and {\it Harvard}. However, it is simple and effective, and --- as our empirical results show --- the resolution is sufficient to establish strong regularities in the timing events.}.

The city-based entity resolution resulted in a dataset with approximately 10\% of the author-paper-pairs being geo-tagged. 
Based on these known geo-locations, we will now fill in the missing ones. 
To do so, we essentially employ {\it Label Propagation}~\cite{bengio06lp,zhu02} (LP), a semi-supervised learning algorithm, to propagate the known cities to the unknown author-paper-pairs based on the similarity between the pairs.
LP works on a graph based formulation of the problem and propagates node labels along the edges. 
We define the LP graph as an undirected graph $G = (V,E)$ with nodes $V$ and edges $E$.
We have a node in $V$ for every author-paper-pair that we want to label with a city.
Every edge $e_{ij} \in E$ between two nodes $i$ and $j$ contains a weight 
$w_{ij}$ that is proportional to the similarity of the nodes.

\begin{figure}[t]
\centering
\scalebox{.8}{
\renewcommand{\arraystretch}{0.7}
\begin{tabular}{lllrr}
\toprule
Id & A & Y & Aff & Aff*\\
\midrule
1 & 1 & 2000 & g & g\\
2 & 2 & 2000 & b & b\\
3 & 2 & 2001 & r & r\\
4 & 1,2 & 2002 & ?,? & r,r\\
5 & 1 & 2002 & ? & r\\
6 & 2 & 2003 & r & r\\
7 & 1 & 2004 & r & r\\
8 & 2 & 2004 & g & g\\
\bottomrule
\end{tabular}
}
\caption{Example database.\label{fig:example_db}}
\end{figure}

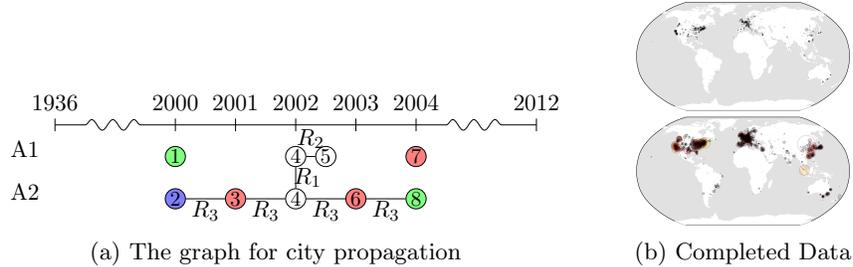
\begin{figure}[t]
\centering
\newsavebox{\tempbox}
\savebox{\tempbox}{
\includegraphics[width=0.23\textwidth]{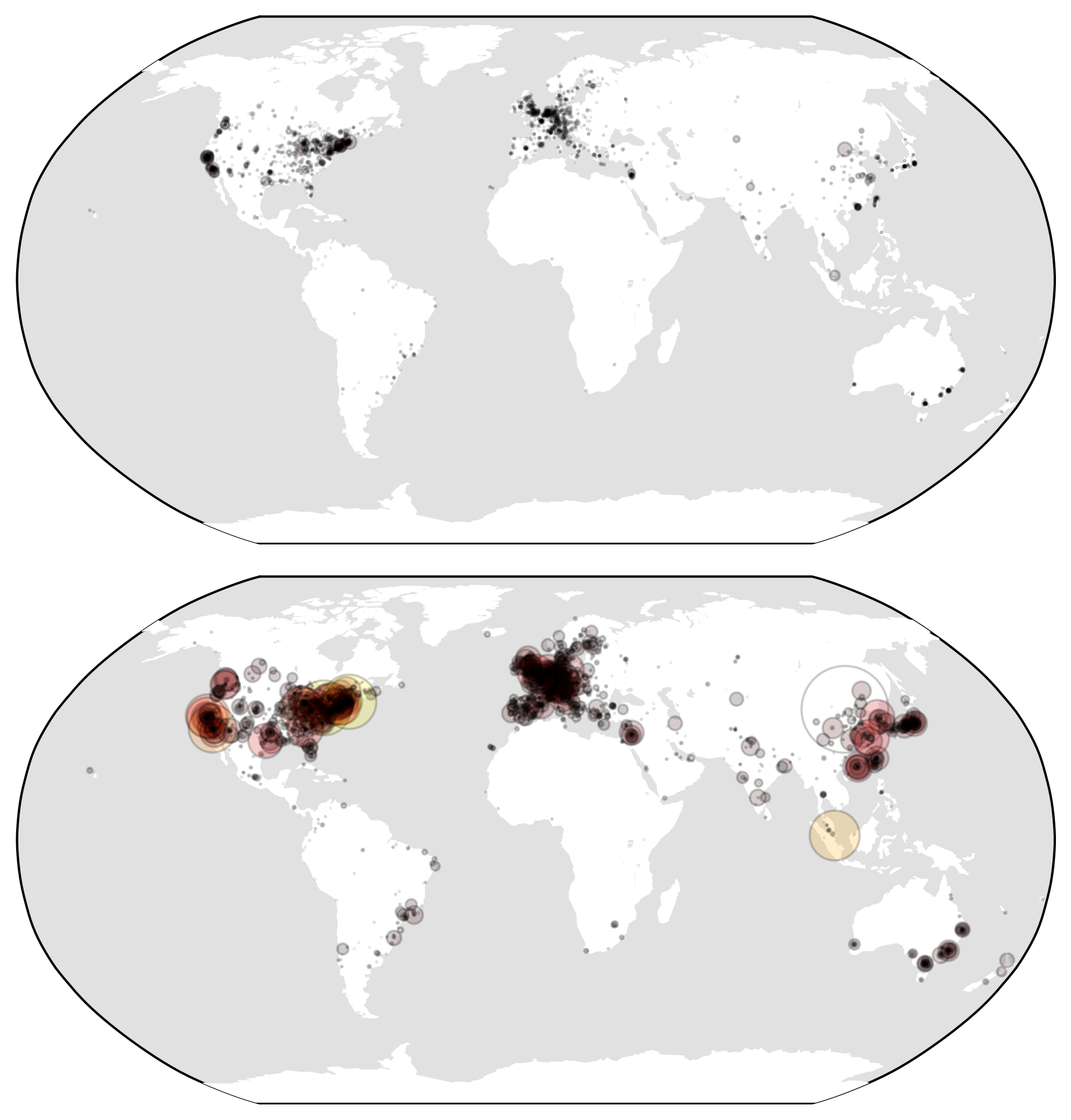}
}
\subfloat[The graph for city propagation\label{fig:graph_construction}]{
         \vbox to \ht\tempbox{%
         \vfil
         \hsize=.60\textwidth
          \scalebox{0.80}{
          \begin{tikzpicture}

\tikzstyle{pub}=[draw=black,circle,inner sep=.5pt]

\draw[decorate,decoration={snake,pre length=5mm, post length=5mm}] (0,0) -- (2,0);
\draw (2,0) -- (6,0);
\draw[decorate,decoration={snake,pre length=5mm, post length=5mm}] (6,0) -- (8,0);

\foreach \x in {0,2,3,4,5,6,8}
   \draw (\x cm,3pt) -- (\x cm,-3pt);

\draw (0,0) node[above=3pt] {$ 1936 $} node[above=3pt] {$  $};
\draw (2,0) node[above=3pt] {$ 2000 $} node[above=3pt] {$  $};
\draw (3,0) node[above=3pt] {$ 2001 $} node[above=3pt] {$  $};
\draw (4,0) node[above=3pt] {$ 2002 $} node[above=3pt] {$  $};
\draw (5,0) node[above=3pt] {$ 2003 $} node[above=3pt] {$  $};
\draw (6,0) node[above=3pt] {$ 2004 $} node[above=3pt] {$  $};
\draw (8,0) node[above=3pt] {$ 2012 $} node[above=3pt] {$  $};

\draw (-0.5,0) node[below=4pt] {A1};
\draw (2,0) node[pub,below=10pt,fill=green!50] (a11) {\small 1};
\draw (4,0) node[pub,below=10pt] (a12) {\small 4};
\draw (4.5,0) node[pub,below=10pt] (a13) {\small 5};
\path (a12) edge node[yshift=3mm] {$R_2$} (a13);
\draw (6,0) node[pub,below=10pt,fill=red!50] (a14) {\small 7};

\draw (-0.5,0) node[below=24pt] {A2};
\draw (2,0) node[pub,below=30pt,fill=blue!50] (a21) {\small 2};
\draw (3,0) node[pub,below=30pt,fill=red!50] (a22) {\small 3};
\draw (4,0) node[pub,below=30pt] (a23) {\small 4};
\draw (5,0) node[pub,below=30pt,fill=red!50] (a24) {\small 6};
\draw (6,0) node[pub,below=30pt,fill=green!50] (a25) {\small 8};
\path (a21) edge node[yshift=-2mm] {$R_3$} (a22)
      (a22) edge node[yshift=-2mm] {$R_3$} (a23)
      (a23) edge node[yshift=-2mm] {$R_3$} (a24)
      (a24) edge node[yshift=-2mm] {$R_3$} (a25);
      
\path (a12) edge node[xshift=2mm] {$R_1$} (a23);
\end{tikzpicture}
}}}
\quad\quad
\subfloat[Completed Data\label{fig:world_map_after_cp}]{%
\vbox{\usebox{\tempbox}\vspace{0pt}\hsize=.24\textwidth}
}
\caption{City Propagation: Missing geo-tags from the example database 
(see~\ref{fig:example_db}) are estimated by propagating the known 
cities/geo-locations across the network of authors and papers. 
The graph for propagating the information (a) is constructed as follows. 
For each author {\it A} and paper {\it Id} there is a node.
Two nodes are connected if they are written by the same author in the same or 
subsequent years or if two researchers co-author them. 
The colors of nodes indicate known cities and white nodes indicate unknown locations.  
As one can see (b), this significantly improves the content of our database. 
The number of geo-tagged author-paper-pairs increased significantly, showing 
the publication activities across the world. (Best viewed in color
\label{fig:author_movement}}
\end{figure}

We will now explain in detail when two nodes are connected by an edge and how the weight $w_{ij}$ for that edge is set.
Intuitively, the weight of an edge is proportional to the similarity of the nodes and we define the similarity of two nodes based on relations such as co-authorship between the authors associated with the nodes.
Only those nodes are connected via an edge where $w_{ij} > 0$. Specifically,
in order to define the edges, we considered the following functions over the set of nodes that 
return facts about the nodes: \texttt{author($i$)}, \texttt{paper($i$)}, and \texttt{year($i$)}.
For example, \texttt{author($i$)} essentially ``returns'' the author of an author-paper node.
Based on these functions, we can now define logic based rules that add a 
rule-specific weight $\lambda_k$ to every matching edge $e_{ij}$.
Initially, we set all weights $w_{ij}$ to zero. The first rule,
\begin{align*}
w_{ij} += \lambda_1 \text{ if } \texttt{paper($i$)} = \texttt{paper($j$)}
\end{align*}
adds a weight between two nodes if the nodes belong to two authors that co-author the paper associated with nodes $i$ and $j$. The second rule,
\begin{align*}
w_{ij} += \lambda_2 \text{ if }  \texttt{author($i$)} = \texttt{author($j$)} \land 
                           \texttt{year($i$)} = \texttt{year($j$)}
\end{align*}
adds a weight whenever two nodes corresponds to different publications by the same author in the same year. Finally, 
\begin{align*}
w_{ij} += \lambda_3 \text{ if }  \texttt{author($i$)} = \texttt{author($j$)} \land
                           \texttt{year($i$)} = \texttt{year($j+1$)}
\end{align*}
fires when the nodes belong to two publications of the same author but written in subsequent years.
This construction process is depicted in Fig.~\ref{fig:graph_construction} for the example publication database in Fig.~\ref{fig:example_db}. 
The example database is missing the affiliation information for papers $4$ and $5$ which is denoted by the ``?'' in the ``Aff''-column.

Based on the constructed graph, we can now build a symmetric $(n \times n)$ similarity matrix $W$ that is used as input to LP. Essentially, LP performs the following matrix-matrix-multiplication until convergence:
$Y^{t+1} = W\cdot Y^{t}\;,$
where $Y^t$ is the labels matrix. In $Y^t$, row $i$ corresponds to a distribution over the possible labels for a node $i$. In $Y^0$, we set a cell $Y_{ij}$ to $1$ if we know that node $i$ has label $j$. All other cells are set to $0$. 
After every iteration, a {\it push-back phase} clamps the rows of the known nodes in $Y^t$ to their original distribution as in $Y^0$.
This operation is performed until convergence or a maximum number of iterations has been reached. 
At convergence, the labels of the unknown nodes are read off the labels matrix, i.e. the label of node $i$ is given by
$y_i = \argmax_{0 \le j \le n-1} Y_{ij}\;.$
In our context, we call this {\bf City Propagation} (CP), that is we run LP on the graph, constructed based on logical rules, to get a distribution over the possible cities for every unlabeled node. 

Although the implementation of CP is just a simple matrix-matrix-multiplication, 
this already becomes challenging with $n$ around five million.
While the similarity matrix $W$ is very sparse, the labels matrix $Y$ becomes denser with every iteration. 
Resulting in an almost pure dense matrix if the graph was completely connected. 
With 4k+ labels, the labels matrix already requires more than 160GB with 64bit float numbers. 
Fortunately, one can easily split the labels matrix into chunks and do the multiplications separately. 
However, we still require an efficient implementation for multiplying a sparse-matrix with a dense-matrix.
We implemented CP with the help of LAMA\footnote{\url{http://www.libama.org/}}, a very efficient linear algebra library.
We ran CP for $100$ iterations and determined the maximizing label for every unlabeled node. 
We used $\lambda_1 = 1$, $\lambda_2 = 3$, and $\lambda_3 = 2$ as weights. 
They had been found using a grid search on a small subset of the data. 
After running CP, GeoDBLP contains \nrLabeledAuthAffs{} geo-tagged author-paper-pairs. 

\begin{figure}[t]
\begin{center} 
\fbox{\includegraphics[width=0.45\textwidth]{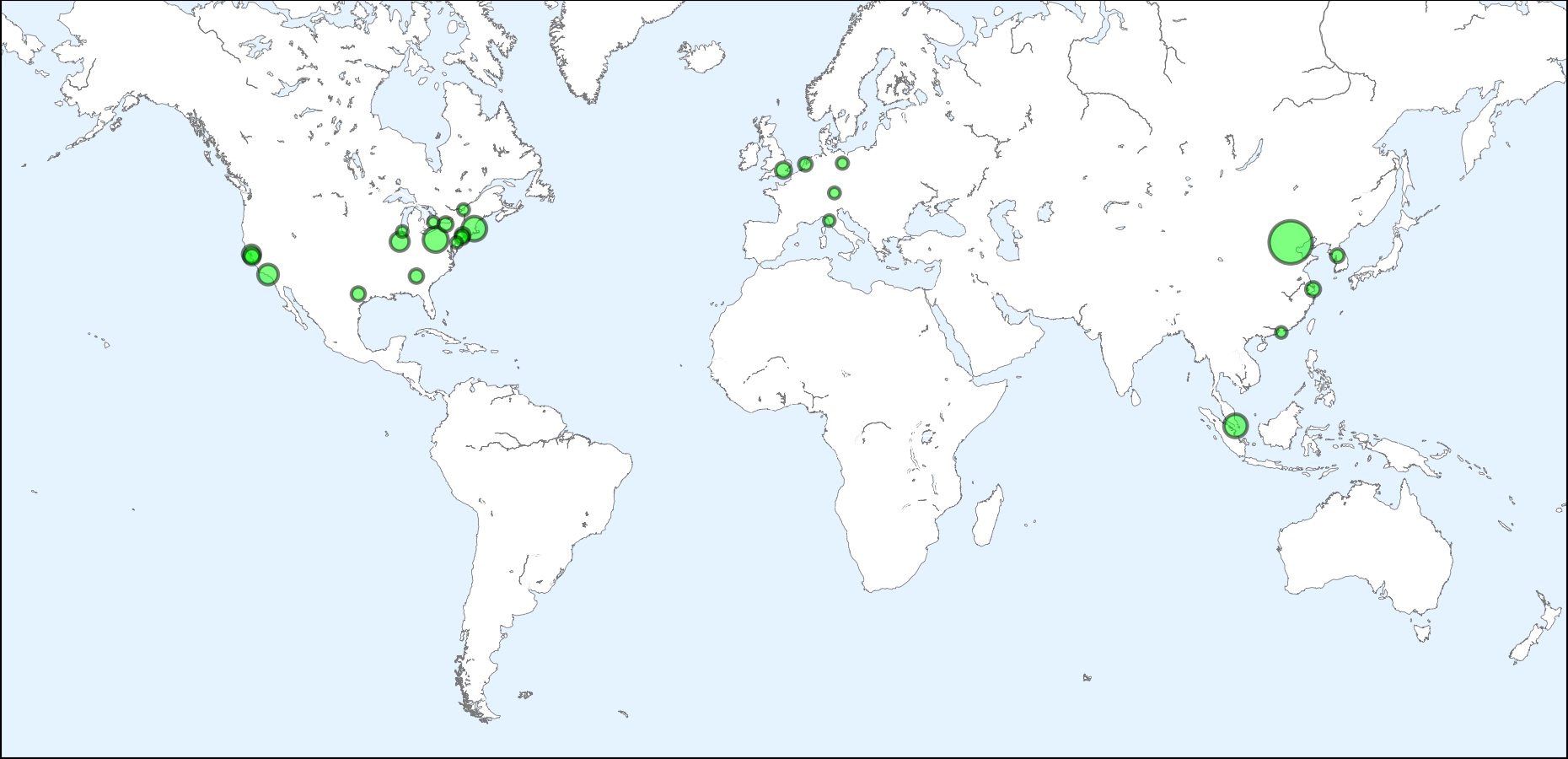}}
\caption{Most productive research cities in the world. The diameter is proportional to the number of publications.\label{fig:nr_pubs}}
\end{center}

\end{figure}


Looking at the last column in our running example in Fig.~\ref{fig:example_db}, 
we see that CP fills the unknown cities, i.e. labels the missing affiliations 
for papers $4$ and $5$. The effect of running CP on our initial dataset is 
shown in Fig.~\ref{fig:world_map_after_cp}. One can see that the worldwide 
productivity increases significantly. The geo-locations of publications alone 
can already reveal interesting insights such as the most productive research 
cities in the world, see Fig.~\ref{fig:nr_pubs}. The main focus of the paper, 
however, is the timing of migration. 

\section{Sketching Migration}
Unfortunately, we cannot directly observe the event of transfer from one residential
location resp. institution to another by a researcher. Instead, we use the affiliations
mentioned in her publication record as a proxy. 
Nevertheless, even after city propagation, 
this list may still be noisy and, hence, does not provide the timing information easily. 
To illustrate this, an author
may very well move to a new affiliation and publish a paper with her old 
affiliation because the work was done while being with the old affiliation. 
Therefore we considered 
{\it migration sketches} only. Intuitively, a sketch captures only the main 
stations of her researcher career. 

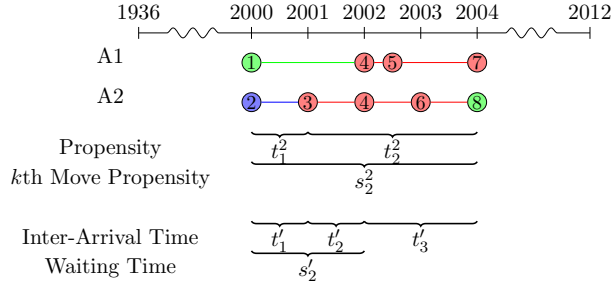
\begin{figure}[t!]
\centering
\scalebox{0.75}{
\begin{tikzpicture}

\tikzstyle{pub}=[draw=black,circle,inner sep=.5pt]

\draw[decorate,decoration={snake,pre length=5mm, post length=5mm}] (0,0) -- (2,0);
\draw (2,0) -- (6,0);
\draw[decorate,decoration={snake,pre length=5mm, post length=5mm}] (6,0) -- (8,0);

\foreach \x in {0,2,3,4,5,6,8}
   \draw (\x cm,3pt) -- (\x cm,-3pt);

\draw (0,0) node[above=3pt] {$ 1936 $} node[above=3pt] {$  $};
\draw (2,0) node[above=3pt] {$ 2000 $} node[above=3pt] {$  $};
\draw (3,0) node[above=3pt] {$ 2001 $} node[above=3pt] {$  $};
\draw (4,0) node[above=3pt] {$ 2002 $} node[above=3pt] {$  $};
\draw (5,0) node[above=3pt] {$ 2003 $} node[above=3pt] {$  $};
\draw (6,0) node[above=3pt] {$ 2004 $} node[above=3pt] {$  $};
\draw (8,0) node[above=3pt] {$ 2012 $} node[above=3pt] {$  $};

\draw (-0.5,0) node[below=4pt] {A1};
\draw (2,0) node[pub,below=10pt,fill=green!50] (a11) {\small 1};
\draw (4,0) node[pub,below=10pt,fill=red!50] (a12) {\small 4};
\draw[color=green, below=10pt] (a11) -- (a12);
\draw (4.5,0) node[pub,below=10pt,fill=red!50] (a13) {\small 5};
\draw (6,0) node[pub,below=10pt,fill=red!50] (a14) {\small 7};
\draw[color=red, below=10pt] (a12) -- (a13) -- (a14);

\draw (-0.5,0) node[below=24pt] {A2};
\draw (2,0) node[pub,below=30pt,fill=blue!50] (a21) {\small 2};
\draw (3,0) node[pub,below=30pt,fill=red!50] (a22) {\small 3};
\draw[color=blue, below=10pt] (a21) -- (a22);
\draw (4,0) node[pub,below=30pt,fill=red!50] (a23) {\small 4};
\draw (5,0) node[pub,below=30pt,fill=red!50] (a24) {\small 6};
\draw (6,0) node[pub,below=30pt,fill=green!50] (a25) {\small 8};
\draw[color=red, below=10pt] (a22) -- (a23) -- (a24) -- (a25);

\draw (-0.5,0) node[below=50pt] {Propensity};
\draw [thick,decoration={brace,mirror},decorate,below=50pt] (2,0) -- (3,0) 
node [pos=0.5,anchor=north] {$t^2_1$}; 
\draw [thick,decoration={brace,mirror},decorate,below=50pt] (3,0) -- (6,0) 
node [pos=0.5,anchor=north] {$t^2_2$}; 

\draw (-0.5,0) node[below=65pt, align=center] {$k$th Move Propensity};
\draw [thick,decoration={brace,mirror},decorate,below=65pt] (2,0) -- (6,0) 
node [pos=0.5,anchor=north] {$s^2_2$};

\draw (-0.5,0) node[below=95pt] {Inter-Arrival Time};
\draw [thick,decoration={brace,mirror},decorate,below=95pt] (2,0) -- (3,0) 
node [pos=0.5,anchor=north] {$t'_1$};
\draw [thick,decoration={brace,mirror},decorate,below=95pt] (3,0) -- (4,0) 
node [pos=0.5,anchor=north] {$t'_2$};
\draw [thick,decoration={brace,mirror},decorate,below=95pt] (4,0) -- (6,0) 
node [pos=0.5,anchor=north] {$t'_3$}; 

\draw (-0.5,0) node[below=110pt, align=center] {Waiting Time};
\draw [thick,decoration={brace,mirror},decorate,below=110pt] (2,0) -- (4,0) 
node [pos=0.5,anchor=north] {$s'_2$};

\end{tikzpicture}
}
\caption{Individual propensities and (inter-)arrival times illustrated for the two researchers A1 and A2 of our running example. A researchers's propensity (shown only for A2) is her probability of migrating. The $k$th move propensities are her probability of making $k>1$ moves. This should not be confused with the (inter-)arrival times of the job market, i.e., of the overall Poisson process. Every node denotes a publication and the node colors denote different affiliations,
i.e. there are three affiliations here: green, red, and blue. From this, we can read off migration: A1 moves from green to red, A2 moves from blue to red and from red to green. (Best viewed in color)\label{fig:hops}}
\end{figure}

\begin{figure}[t]
\centering
\subfloat[Number of Moves\label{fig:hops_per_year}]{
\includegraphics[width=.35\columnwidth]{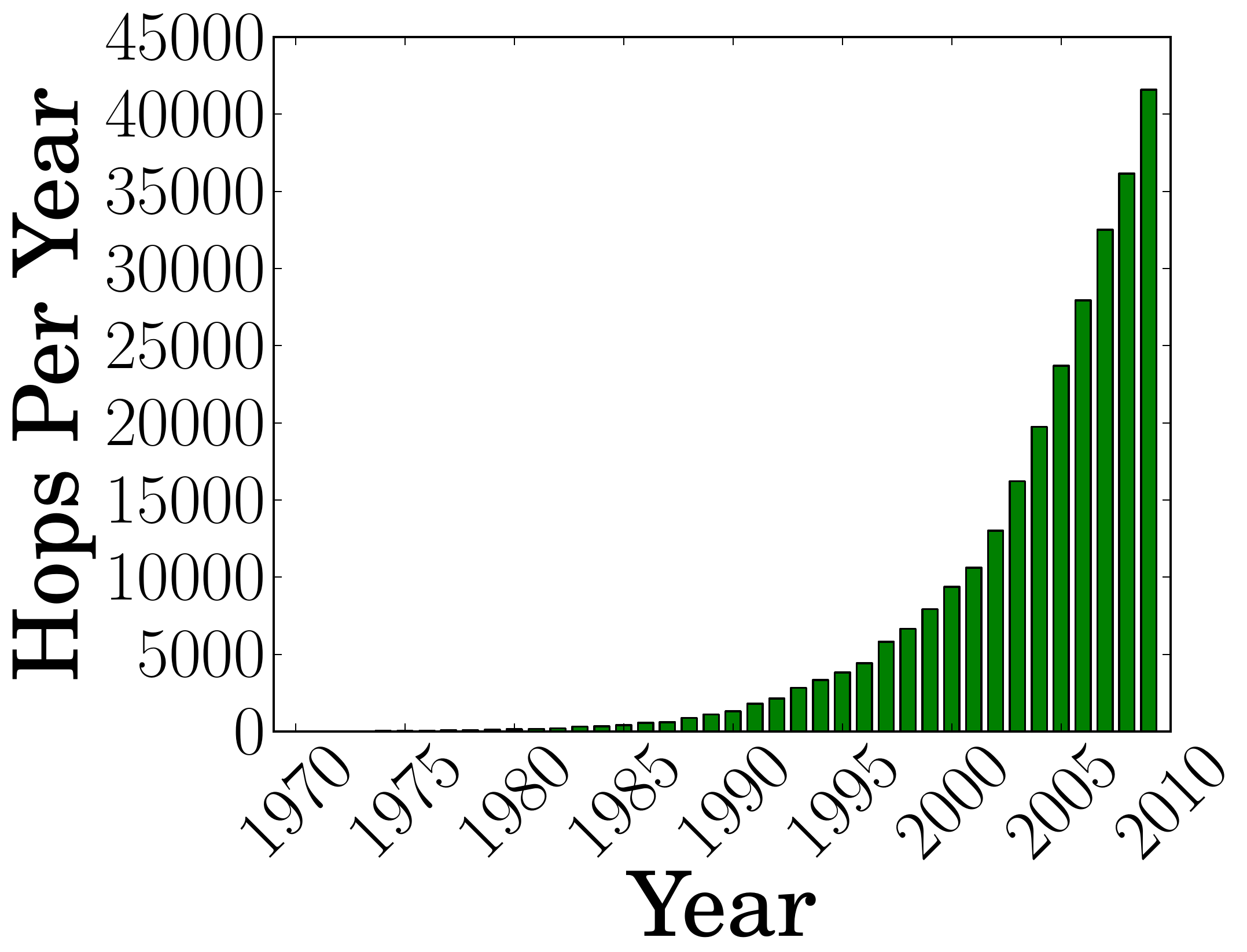}
}
\qquad\qquad
\subfloat[Ratio\label{fig:ratio_hops_authors}]{
\includegraphics[width=.35\columnwidth]{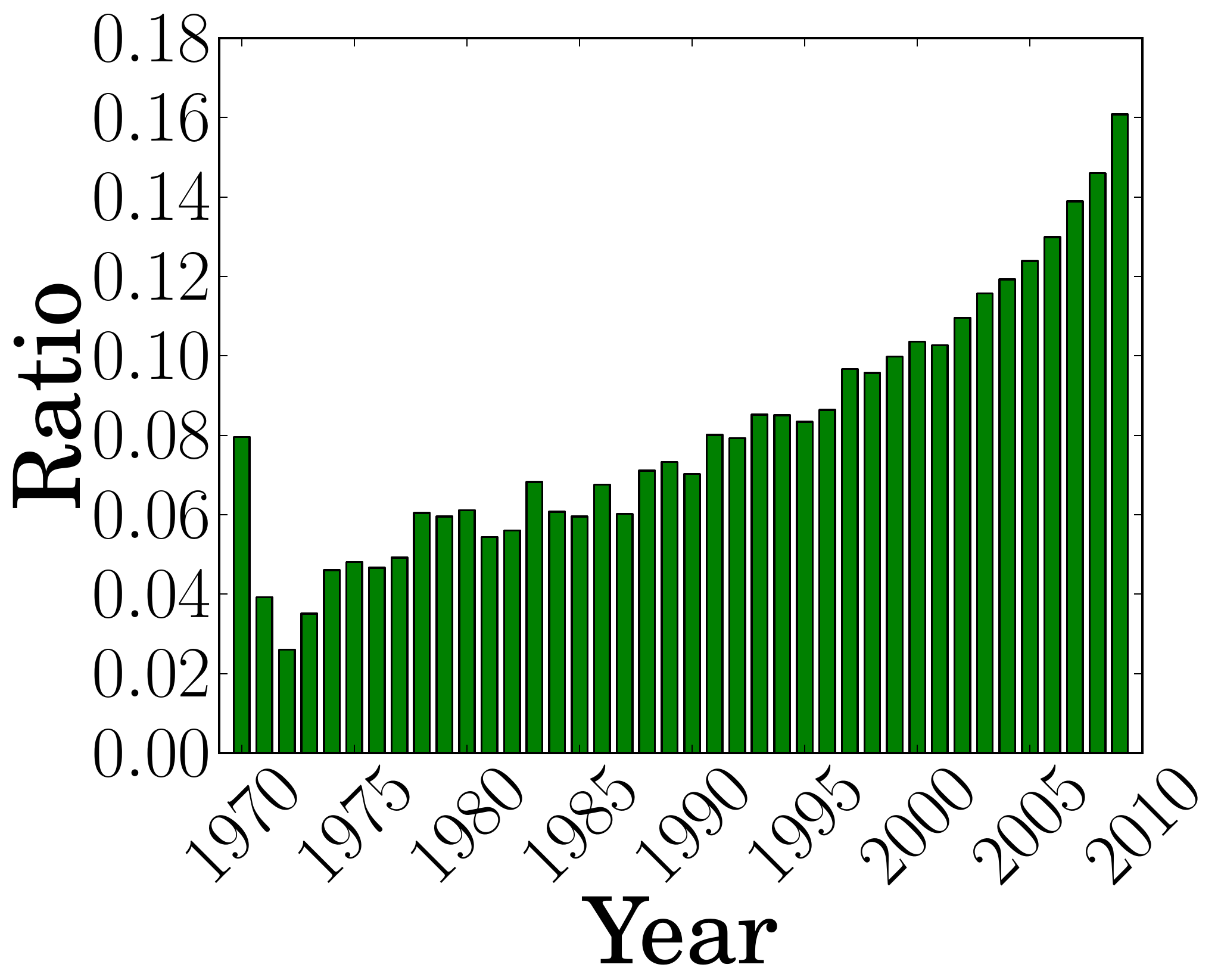}
}
\caption{Migration statistics over time in GeoDBLP. The ration between moves and authors per year  (Fig.~\ref{fig:ratio_hops_authors}) does not grow as fast as the number of hops (Fig.~\ref{fig:hops_per_year}) or authors  (Fig.~\ref{fig:dblp_authors}). Moreover, this illustrates that the job market is actually an inhomogeneous Poisson process that locally, say for periods of 10 years, can well be assumed to be homogeneous.\label{fig:hop_stats}}
\end{figure}

More formally, we define a migration sketch as the set of the unique affiliations of an author ordered
by the first appearance in the list of publications. For instance, in our running example, we
have $[2000: \text{Aff}_g, 2002: \text{Aff}_r]$ for author $\text{A}_1$ and 
$[2000: \text{Aff}_b, 2001: \text{Aff}_r, 2004: \text{Aff}_g]$ for author $\text{A}_2$. That is, 
Author $A_1$ has two different affiliations, $\mbox{\em Aff}_g$ appearing in 2000 the first time and the first publication with $\mbox{\em Aff}_r$ in 2002.
Of course, this approach has the drawback that we can not capture if a person returns to an earlier affiliation after several years. 
Finally, we dropped implausible entries from the resulting sketch database.
For instance, we dropped sketches with more than ten affiliations. 
It is very unlikely that a single person has moved more than ten times and these sketches should rather be attributed to an insufficient entity disambiguation.
Having the migration sketches at hand, we can now define a {\it migration}/{\it move} of a researcher as 
the event of transfer from one residential
location to another by a researcher in her migration sketch. Fig.~\ref{fig:hops} shows the moves 
of author $A_2$ in our running example. In total, we found \nrHops{} migrations in GeoDBLP. 
The number of moves per year is shown in Fig.~\ref{fig:hops_per_year} and it shows 
that the number of moves increases with the years super-linearly. 
However, when we normalize the numbers of moves by the number of scientists, we see roughly a linear slop, see Fig.~\ref{fig:ratio_hops_authors}. 
With this information at hand, we can now start to investigate the statistical properties of researcher migration. 


\section{Regularities of Timing Events}
As mentioned above, reasons to migrate are manifold. Despite this
complex web of interactions, we now show that researcher migration
shows remarkably simple but strong global regularities in the
timing.

\subsection{(R1) Migration Propensity is Log-Normal}
Given the migration sketches, we can now read off timing information. 
First, we estimate the propensity to transfer to a new residential location or institution
across scientists. To do so, let $T^j_i$ be the point in time when a researcher moves from 
one location to the next one. Let $t^j_i$ be the time between the $T^j_{i-1}$ and $T^j_i$. 
We call $t^j_i$, i.e. the time between two moves, the {\it migration propensity} (see Fig.~\ref{fig:hops}). 
It reflects the bias of researchers to stay for a specific amount of time until moving on.

\begin{figure}[t]
\begin{center}
\includegraphics[width=0.35\textwidth,page=3]{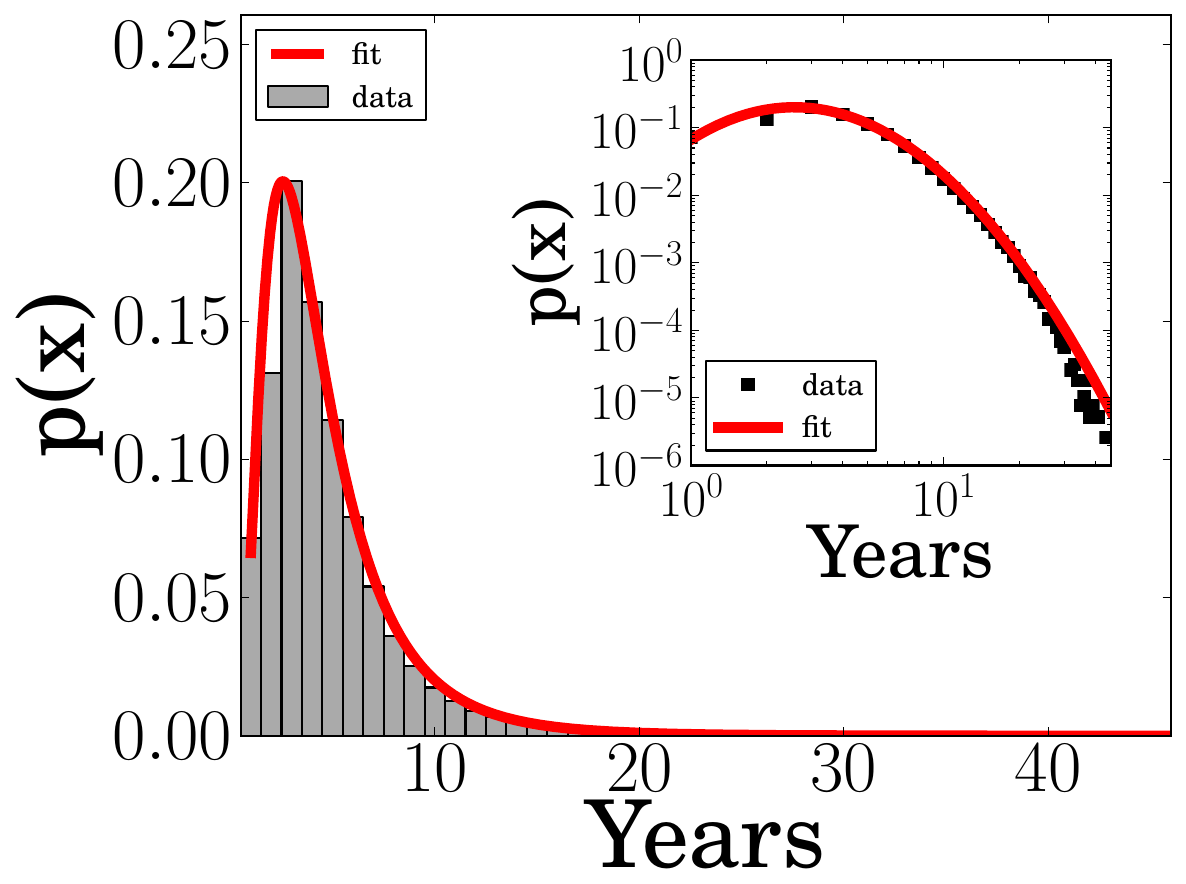}
\end{center}
\caption{Migration propensity. The individual migration propensity  
is best fitted by a log-normal distribution. That is, although jobs enter the market all the time, researchers are generally not ``memoryless'' but have to
care greatly about their next move, and this timing 
is a multiplicative function of many independently distributed factors.
\label{fig:timetohop}}
\end{figure}

\begin{figure}[t]
\centering
\includegraphics[width=0.90\textwidth,clip=true,trim=2.6cm 4.5cm 2.6cm 4.5cm]{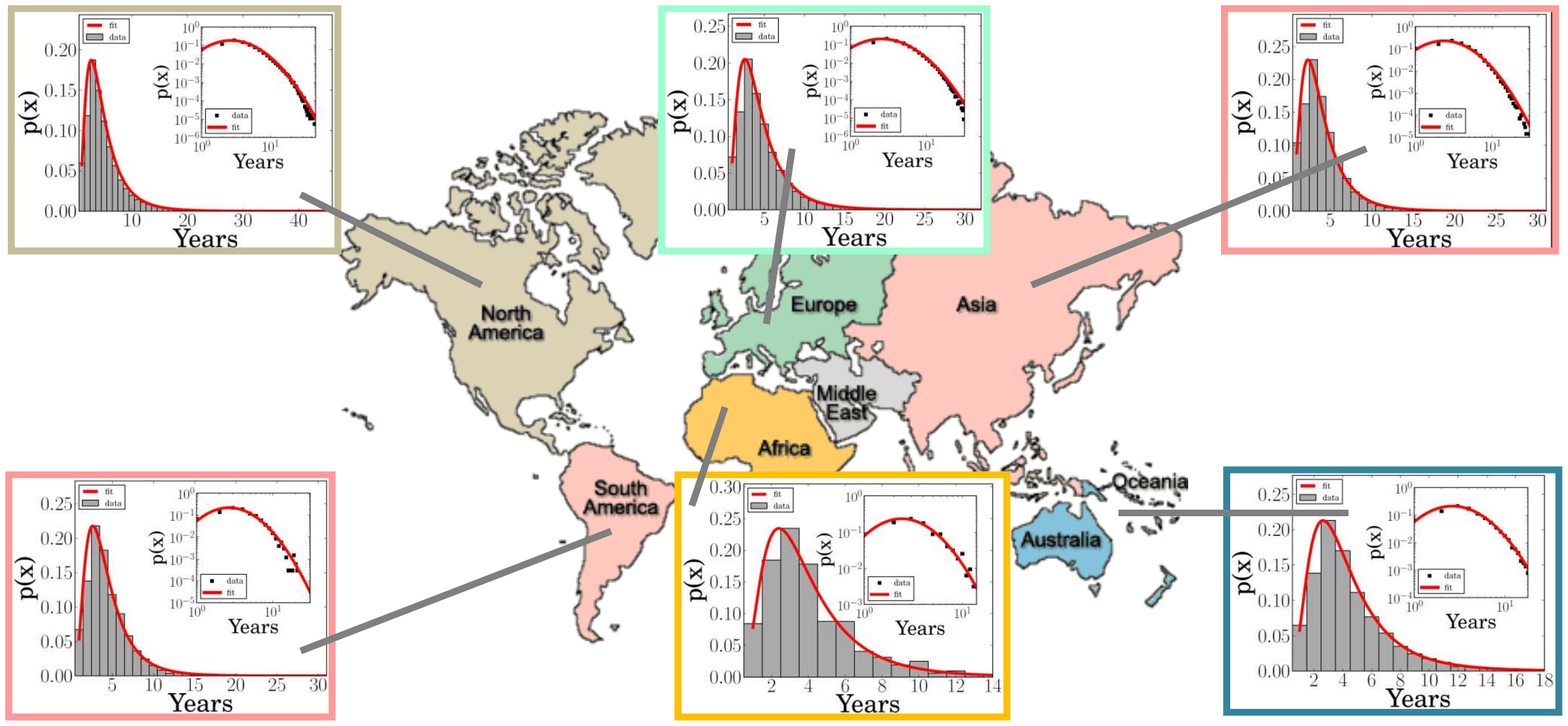}
\caption{Migration propensities are remarkably similar across continents and again best fitted by log-normal. Thus, timing research careers has no cultural boundaries across continents. (Best viewed in color)\label{fig:continents}}
\end{figure}

Fig.~\ref{fig:timetohop} shows the best fitting distribution in terms of log-likelihood and
KL-divergence among various distributions such as log-normal, gamma, exponential, 
inverse-Gauss, and power-law using maximum likelihood estimation for the parameters. 
It is a log-normal distribution~\cite{aitchison57lognormal,stewart94info}. 
That is, the log of the propensity is normal distribution with density
\begin{equation}
\mbox{\em ln}(x) = \frac{1}{x\sqrt{2\pi\sigma^2}}e^{-\frac{(\ln x - \mu)^2}{2\sigma^2}}\;.
\end{equation}
The parameters $\mu$ and $\sigma^2 > 0$ are the mean and the standard deviation of the variable's natural logarithm.
This is a plausible model due to Gibrat's ``law of proportionate effects''~\cite{gibrat30}. The underlying
propensity to move is a multiplicative function of many independently distributed
factors, such as motivation, open positions, short-term contracts, among others. That is,
such factors do not add together but are multiplied together, as a weakness in
any one factor reduces the effects of all the other factors. 
That this leads to log-normality can be seen as follows.
Recall that, by the law of large numbers, the sum of independent random variables becomes a normal distribution regardless of the distribution of the individuals. 
Since log-normal random variables are transformed to normal random variables by taking the logarithm, when random variables are multiplied, as the sample size increases, the distribution of the product becomes a log-normal distribution regardless of the distribution of the individuals. This might explain why the log-normal 
distribution is one of the most frequently observed distributions in nature 
and describes a large number of physical, biological
and even sociological phenomena~\cite{limpert01bioscience}. For example, variations
in animal and plant species just as in incomes appear log-normal, i.e. normal when presented
as a function of logarithm of the variable. Dose-response relations just as grain sizes from grinding
processes show log-normal distributions.  Moreover, although the overall job market is a 
Poisson process, as we will show later on, 
it is good that the migration propensity is not exponential. It is precisely this 
non-Poisson that makes it possible to make predictions based on past observations.
Since positions are occupied in a rather regularly way, upon taking a position it is very unlikely that you will take up another position soon. 
In the Poisson case, which is the dividing case between clustered and regular processes, you should be indifferent to the time since the last position. 

\begin{figure}[t]
\centering
\includegraphics[width=0.7\textwidth,clip=true,trim=1cm 4.5cm 1cm 0.5cm]{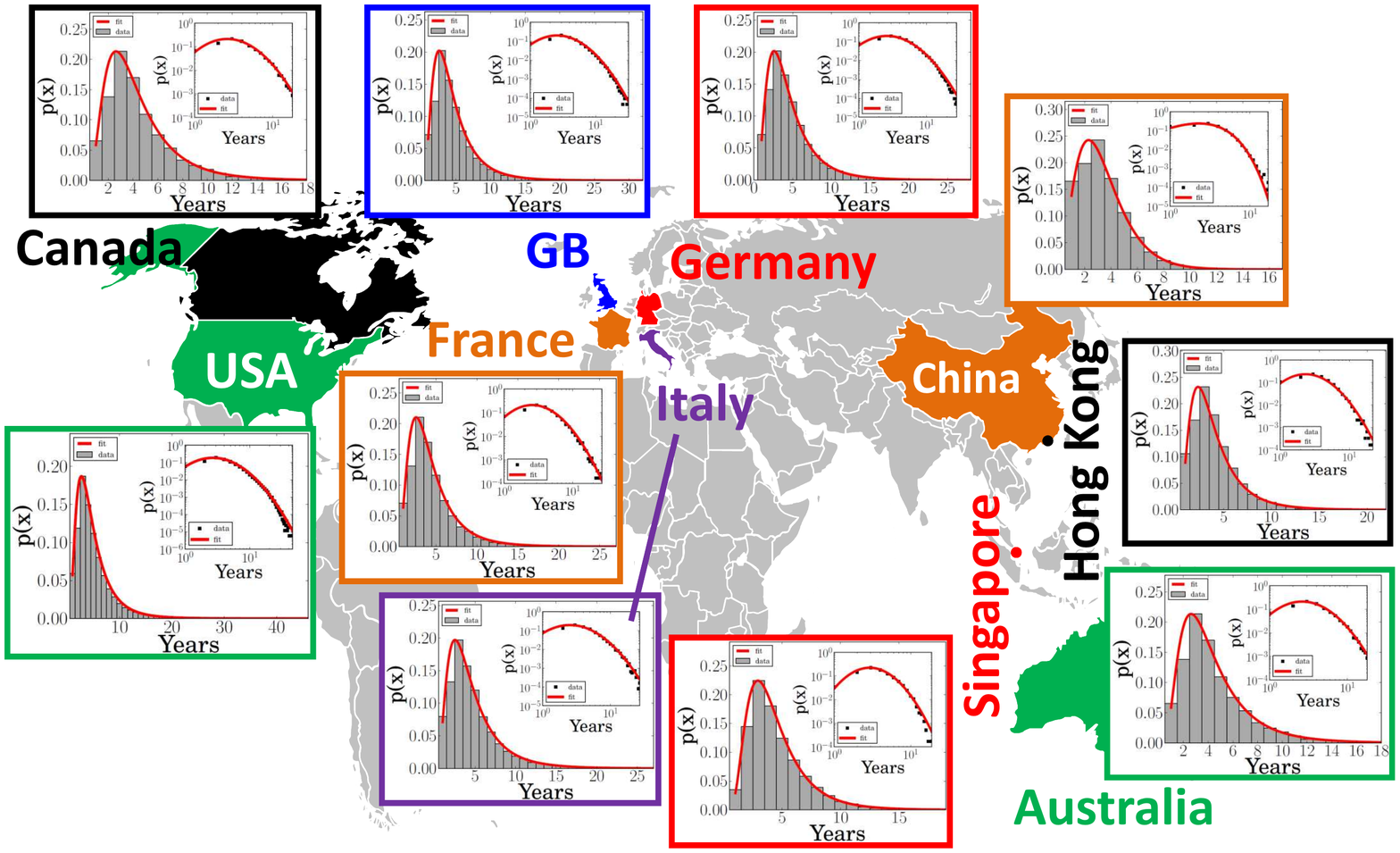}
\caption{Zooming in on migration propensities: across the most productive countries they are best fitted by log-normal. Actually, the representative countries USA, China, Germany, UK, Australia, Singapore, Canada, France, Italy, and Hong Kong are shown. Except for China, all are best fitted by log-normal. China's migration propensity follows a gamma distribution. 
(Best viewed in color)\label{fig:countries}}
\end{figure}

Based on our data, a computer scientist stays on average 5 years at a place. Thus headhunters, for example, should approach young potentials in their fourth year. On the other hand, one should probably reconsider the common practice, e.g. in the EU and the US, of having projects lasting only three years to fill in the gap.
More importantly, the log-normality of the propensity can be found across continents and countries of the world, see Figs.~\ref{fig:continents} and \ref{fig:countries}, where we considered only moves originating from a 
continent resp. country. Timing research careers has clearly no cultural boundaries!

\begin{figure}[t]
\begin{center}
\includegraphics[width=0.35\textwidth,page=2]{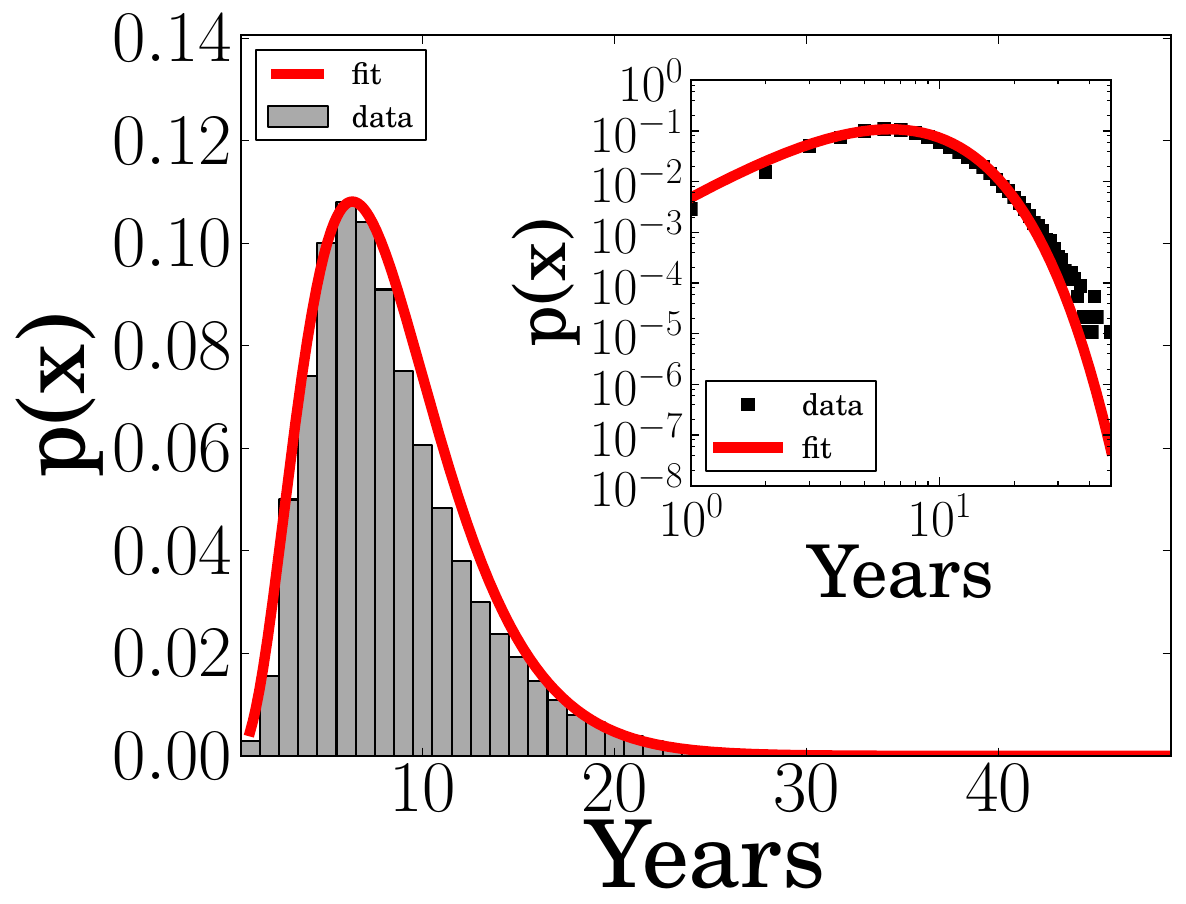}
\qquad\qquad
\includegraphics[width=0.35\textwidth,page=2]{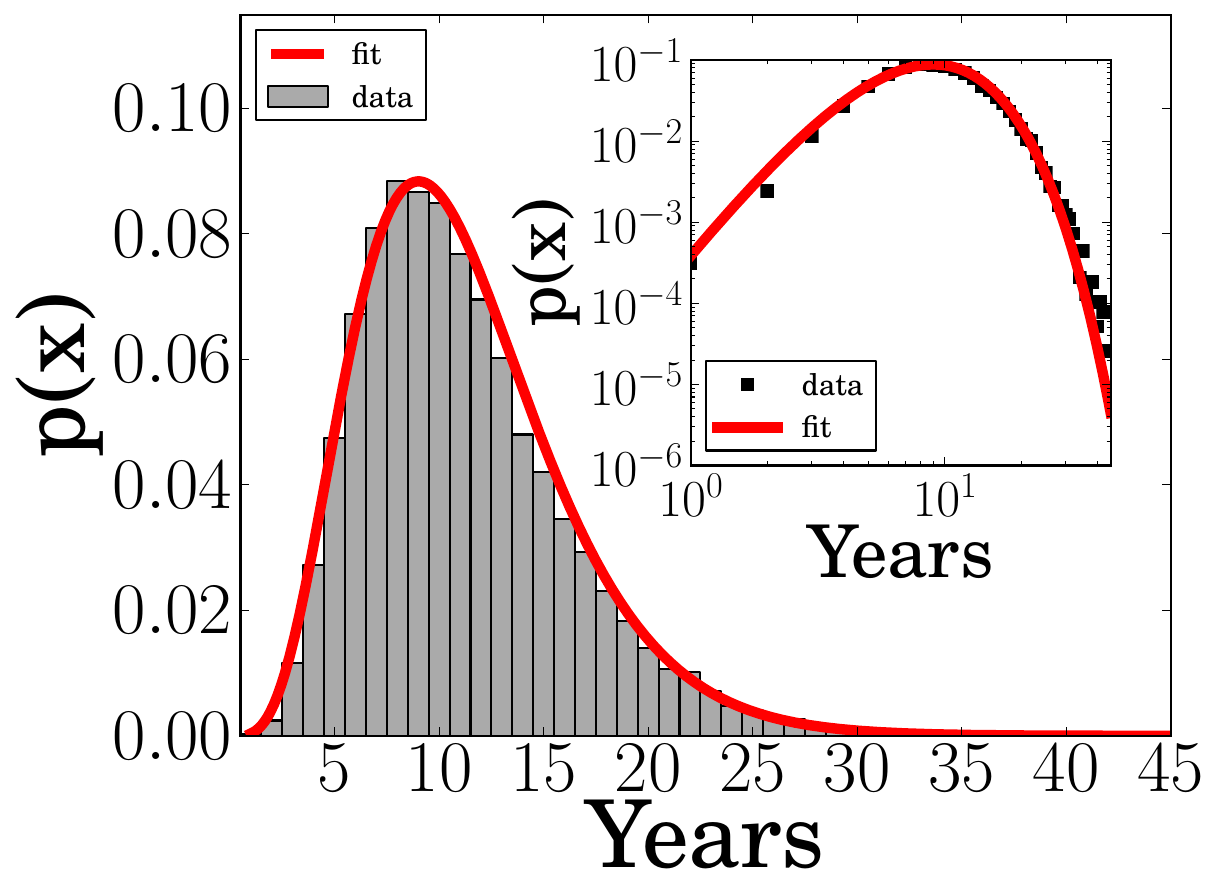}\\
\includegraphics[width=0.35\textwidth,page=2]{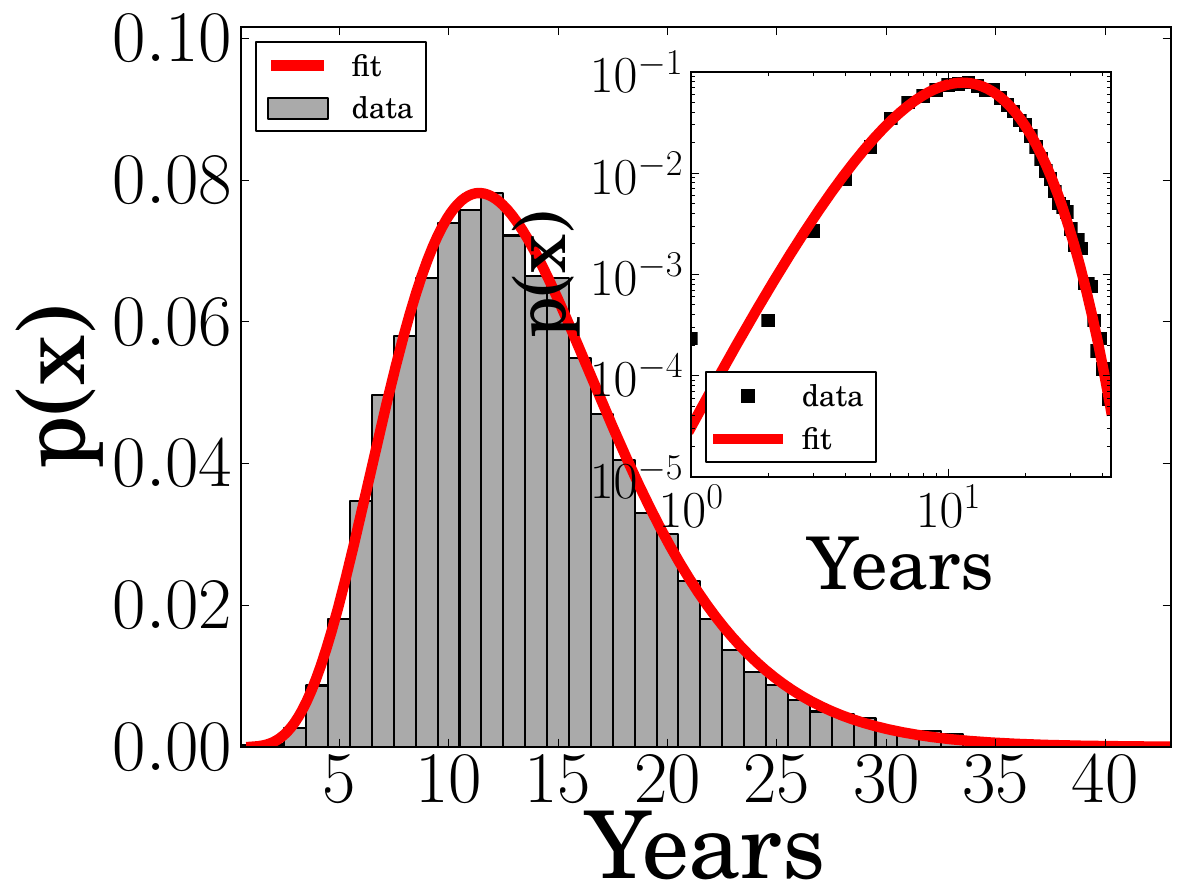}
\qquad\qquad
\includegraphics[width=0.35\textwidth,page=2]{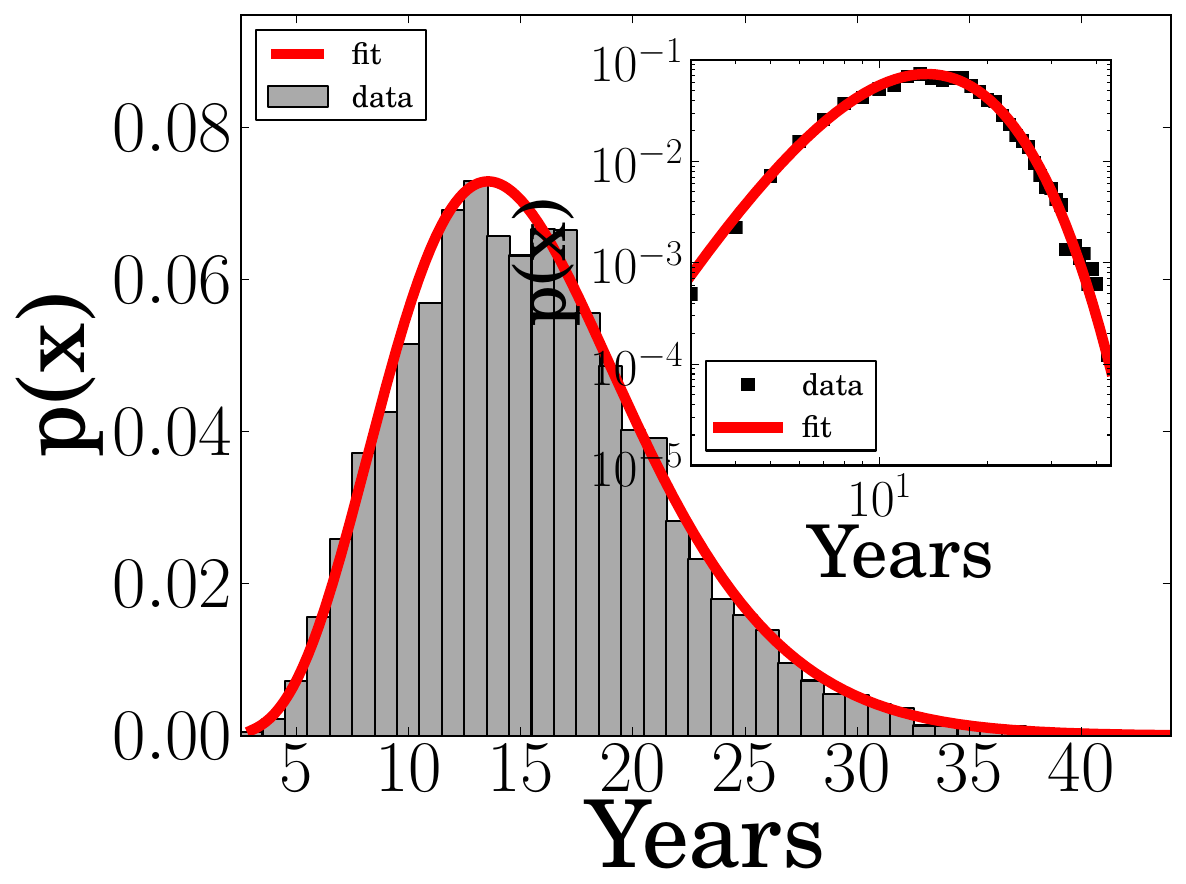}
\end{center}
\caption{$k$th move propensities. 
The $k$th move propensities (left-right, top-down with $k=2,3,4,5$) are best fitted by gamma distributions. 
This suggests that migration at later career stages is ``memoryless'', i.e., 
it follows an exponential distribution.\label{fig:time_to_kth_move}}%
\end{figure}

\subsection{(R2) k-th Move Propensities are Gamma}
Fig.~\ref{fig:time_to_kth_move} shows the best fitting distribution in terms 
of log-likelihood and KL-divergence among various distributions such as 
log-normal, gamma, exponential, inverse-Gauss, and power-law using maximum 
likelihood estimation for the propensity to make $k>1$ migrations. 
More precisely, the $k$th move propensity for an author $A_i$ is defined as $s^i_k = \sum_{j=1}^k t^i_j$.  
It is a
gamma distribution,
\begin{equation}
\mbox{\em ga}(x) = \frac{1}{\Gamma(k)\theta^k} x^{k-1}e^{-\frac{x}{\theta}}\;,
\end{equation} 
with shape $k>0$, scale $\theta > 0$, and $\Gamma(k) = \int^\infty_0 s^{k-1}e^{-s} ds$\;,
suggesting that migration at later
career stages is ``memoryless''. Why? Well, this follows from the theory of Poisson processes.
For Poisson processes, we know that the inter-arrival times are independent and obey an exponential form,
$\mbox{\em exp}(t) = \lambda e^{-\lambda\cdot t}\;$, 
where $\lambda > 0$ is called the intensity rate.
The important consequence of this is that the distribution
of $t$ conditioned on $\{t > s\}$ is again exponential. That is, the remaining time 
after we have not moved to a new position at time $s$ has the same distribution as the 
original time t, i.e., it is memoryless. Moreover, we know that the time until 
the $k$-th move --- the $k$th move propensity --- has a gamma distribution; it is the sum of the first $k$ 
propensities of senior researchers. So, the propensities for the next move 
turn exponential for later career stages. 
This is plausible, since early career researchers 
have seldom taken many positions and, hence, we consider here rather senior researchers,
which typically have permanent positions; they do not have to greatly care about their moves. As a consequence, e.g. competing universities have to top the current
position of a senior researcher if they want to hire her.

\subsection{Job Market is Poisson Log-Normal}
So far, we have shown that the propensities, let us call it $\delta$, to move 
to a new residential location resp. institute follow a log-normal distribution. 
We have also shown that $k$th move propensities
follow a gamma distribution, suggesting that  
propensities of senior researchers are exponential. The latter fact already points towards a
Poisson model. More precisely, we postulate that the job market follows a Poisson-log-normal
model~\cite{stewart94info}. That is, given a specific scientist's migration propensity $\delta$,
her probability of migrating follows a simple Poisson model:
$\mbox{\em pos}(k) = 1\slash{k!}\cdot(\delta^ke^{-k})\;$,
for $k=1,2,3,4,...$. Thus the rate of the Poisson process is a function of the migration propensity. 
The number of migrations 
for all scientists having the same $\delta$ value will follow the same Poisson process.
Moreover, since the sum of Poisson processes is again a Poisson process, we know that 
every finite sample of scientists with $\delta$s drawn from a log-normal is again
following a Poisson process. Thus, assuming the job market to be a Poisson model 
is plausible. It actually tells us that the arrival of job openings is memoryless. 
Open positions should always be announced as they come. On a global scale, 
there is no point in waiting to announce them. 
There are always researchers ready to take it. And, individual researchers can 
always look out for new job openings. 

\begin{figure}[t]
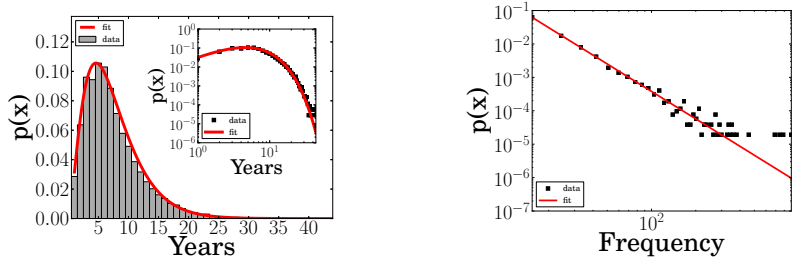

\centering
\includegraphics[width=0.35\textwidth,page=2]{figs/sketches_time_to_return}
\qquad\qquad
\includegraphics[width=0.35\textwidth]{figs/sketches_connections_gt10_log}
\caption{(Left) Brain circulation follows a gamma distribution. (Right) The inter-city migration frequency follows a power-law (after removing low-frequency connections).\label{fig:time_freq}}
\end{figure}
\subsection{(R3) Brain Circulation is gamma}
Brain circulation, or more widely known as brain drain, is the term generically
used to describe the mobility of high-level personnel. It is an emerging global
phenomenon of significant proportion as it affects the socio-economic and socio-cultural
progress of a society and a nation, and the world. Here, we defined it as the time 
until a researcher returns to the country of her first publication. Only 
$\nrAuthorsReturned{}$ out of $\nrMobileAuthors{}$ ($15\%$) mobile researchers, 
i.e., researchers that have moved at least once, and out of a total of 
$1,080,958$ ($3\%$) researchers returned to their roots (in terms of publications).
As to be expected from the statistical regularity for $k$th move propensities, it 
also follows a gamma distribution, as shown in Fig.~\ref{fig:time_freq}(left). 
Since a gamma distribution is the sum of exponential distributions, returning is
memory less. Researchers cannot plan to return to their roots but
rather have to pick up opportunities as they arrive.

\section{Link Analysis of Migration}
Link analysis techniques provide an interesting alternative view on our migration data. That is, we view 
migration as a graph where nodes are cities and directed edges are migration links between cities.
More formally, the author-migration graph is a directed graph $G = (V,E)$ where each vertex 
$v \in V$ corresponds to a city in our database.  There is an edge $e \in E$ from vertex $v_1$ 
to vertex $v_2$ iff there is an author who has moved from an affiliation in city $v_1$ to $v_2$.

%


\subsection{(R4) Inter-City Migration is Power-Law}
Triggered by Zipf's early work and other recent work on inter-city migration~\cite{zipf46asr,cohen08pnas, Siminil12nature}, we investigated the frequency of inter-city researcher migration. 
The frequency of a connection between two cities can be seen as knowledge exchange rate between the cities. 
It is a kind of knowledge flow because one can assume that researchers take their acquired knowledge to next affiliation.
If one looks at the author-movement-graph as a traffic network, high frequent connections corresponds to highly used streets.
Fig.~\ref{fig:time_freq}(right) shows the distribution with a fitted power-law using maximum likelihood estimation. A likelihood comparison to other distributions such as log-normal and gamma revealed that 
a power-law is the best fit. Thus, there are only few pairs of cities with frequent researcher exchange 
and many low-frequent pairs. However, cities with a high exchange of researchers
will exchange even more researchers in the future. Investments into migration pay off.

\subsection{(SP 1) Migration Authorities and Hubs}
Next, we are interested in mining the migration authorities and hubs. 
To do so, we
use Kleinberg's HITS-algorithm\cite{kleinberg99} on the author-migration graph. 
The algorithm is an iterative power method and returns two scores for every node 
in the graph, which are known as {\it hubs} and {\it authorities}.
This terminology arises from the web where hubs and authorities represent websites.
Hubs are pages with many outlinks and authorities are pages with many inlinks.

\begin{figure}[t]
\centering
\fbox{\includegraphics[height=0.18\textwidth]{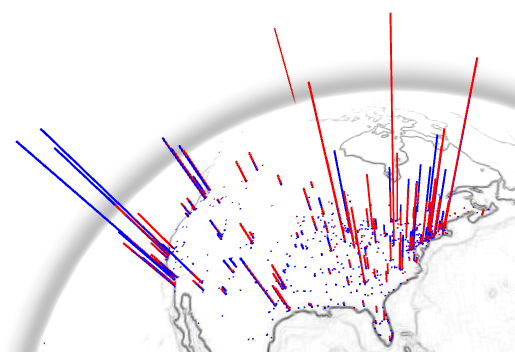}}
\quad
\fbox{\includegraphics[height=0.18\textwidth]{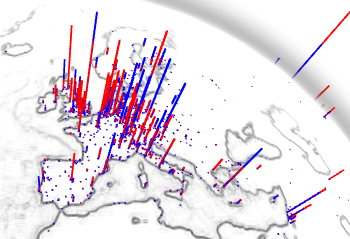}}
\quad
\fbox{\includegraphics[height=0.18\textwidth]{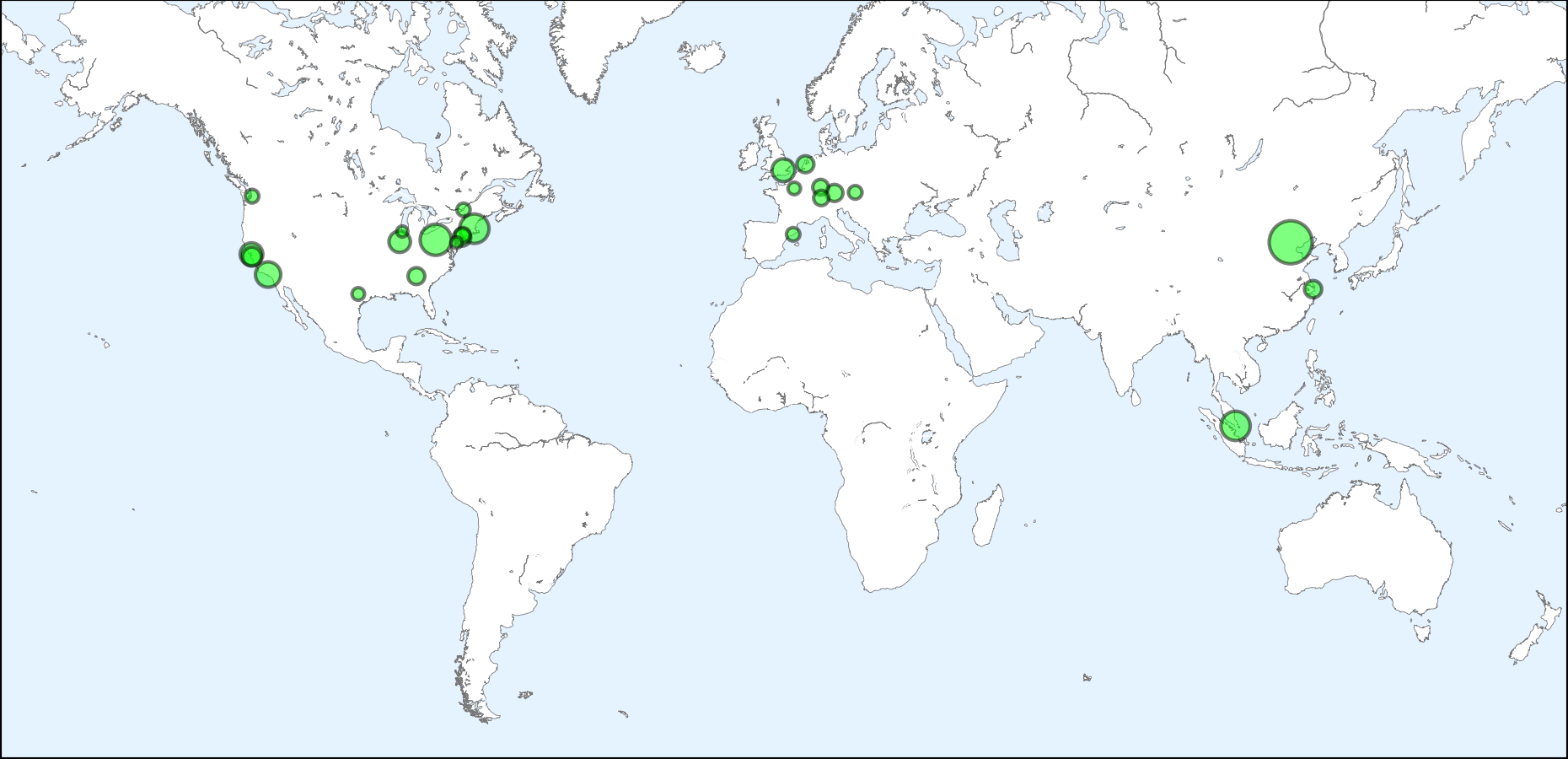}}
\caption{(left and middle) Running HITS on the directed author-migration graph reveals sending, receiving, and incubator countries. Shown are representative cities in North America (left) and Europe (middle). The size of spikes encodes the value of the ``authority'' (blue) and ``hub'' (red) scores. Incubator cities have well balanced scores. As one can see, the European cities rather send  researchers. US cities at the east cost are incubators, and west cost cities receive researchers.\label{fig:hits}
(right) Top 25 migration cities ranked by PageRank. Compared to the productivity map in Fig.~\ref{fig:nr_pubs}, one can see that productive cities are not necessarily cities with high migration flux. (Best viewed in color)\label{fig:pr}
}
\end{figure}

In our context, inlinks correspond to researchers arriving in a city --- she picks up a new position --- whereas an outlink 
corresponds to a researcher leaving a city --- e.g. funding ends.
Hubs can be seen as ``sending'' cities, i.e., they send out researcher across the world. 
On the other hand, authorities can either be cities where people want to stay and tenure positions are available or where people drop out of research, i.e. heading to industry. They are 
``receiving'' cities. Moreover, if we make the assumption that only high-quality students and 
scientists get new positions, one may view sending cities as institutions producing high profile 
scientists but also cannot hold all of them, due to restricted capacities or low attractiveness.
In contrast, receiving cities might have the capacities and reputation to hold many migrating
researchers or highly interesting industrial jobs are close by. Cities having generally high scores are incubators: they attract a lot of migration but also send them to other places. 

Fig.~\ref{fig:hits} shows the sending and receiving scores for cities in the representative 
regions of the US and Europe\footnote{Rendered with WebGL Globe (see \url{http://www.chromeexperiments.com/globe}).}. The US clearly shows an East-coast/west-coast movement. 
The east coast aggregates many sending cities while receiving cities dominate the west coast.
This is plausible. Not only are there many highly productive universities on the west coast, see also Fig.~\ref{fig:nr_pubs}, but labor market for high-tech workers in, say, the Bay Area is the strongest in a decade. Thousands of new positions are being offered by small startups and established tech giants.
However, one should view many of the east-coast cities as incubators since they have high overall scores. The scores of European cities are typically much smaller, see again
Fig.~\ref{fig:hits}(right). Europe is dominated by sending cities. Few exceptions are Berlin, Munich, Stockholm, and Zurich. The largest receiving city in the world is Singapore. This is also plausible.
The city-state is known for its remarkable investment in research in recent years, as e.g. noted by a recent Nature Editorial~\cite{editorial10nature}\footnote{The recent economic pressure mounting on research communities in Singapore and around the world is not well captured in our data, which lasts to 2010 only.}.
In contrast, the largest sending city by far is Beijing. This is also plausible. There has been upsurge in Chinese emigration to Western countries since the mid-first decade of the 21st century~\cite{lam10}. 
In 2007, China became the biggest worldwide contributor of emigrants.

\subsection{(SP2) Moving Cities}
Following up on HITS, we also computed PageRank on the author-migration graph. 
Compared to HITS, PageRank~\cite{Page99} produces only a single score: a page is 
informative or important if other important pages point to it. More formally, 
by converting a graph to an ergodic Markov chain, the PageRank of
a node $v$ is the (limit) stationary probability that a
random walker is at $v$. In the context of migration, this has a natural and very appealing
analogy. The PageRank computes the (limit) stationary probability that a
random migrator is at a city. 
 
To compute the {\it MigrationRank} of a city, the author-migration graph is transformed 
into the PageRank-matrix on which a power method is applied to obtain the PageRank-vector, 
containing a score for every node in the graph.
The transformed matrix also contains the stochastic adjustment identical to the random surfer in the original work. That is, a researcher can always migrate from one affiliation to another affiliation, 
even if no one else did so before.
Fig.~\ref{fig:pr}(right) shows the top 25 cities in the world according to the MigrationRank.
Compared to the productive map in Fig.~\ref{fig:nr_pubs}, one can clearly see many similarities but 
although notable differences. The US is not only productive but thrives on migration.
Vancouver, B.C., is among the top 25 when it comes to migration but not when it comes to productivity. 
Generally, productivity does not imply a high migration rank. 
Beijing, however, is top in both when it comes to productivity and migration. Singapore is higher ranked for migration
than for productivity. 
European cities seem to also thrive on migration more than on productivity. 
At least there are much more cities in the top 25 than for productivity. However, compared to the US,
they are less clustered together

\section{Conclusions}
International mobility among researchers not only benefits the individual development of scientists,
but also creates opportunities for intellectually productive encounters, enriching
science in its entirety, preparing it for the global scientific challenges lying ahead. 
Moreover, mobile scientists act as ambassadors
for their home country and, after their return, also for their former host
country, giving mobility a culture-political dimension. So far, however, no statistical
regularities have been established for the timing of migration. In this paper, 
we have established the first set of statistical regularities and patterns for research migration 
stemming from inferring and analyzing a large-scale, geo-tagged dataset from the web representing
the migration of all researchers listed in DBLP. The methods and findings highlight the value of
using the World Wide Web together with data mining to fill in missing data
as a world-wide lens onto research migration.

\begin{figure}[t]
\begin{center}
\includegraphics[width=0.45\textwidth,clip=true,trim=1.3cm 7.0cm 1.3cm 2.0cm]{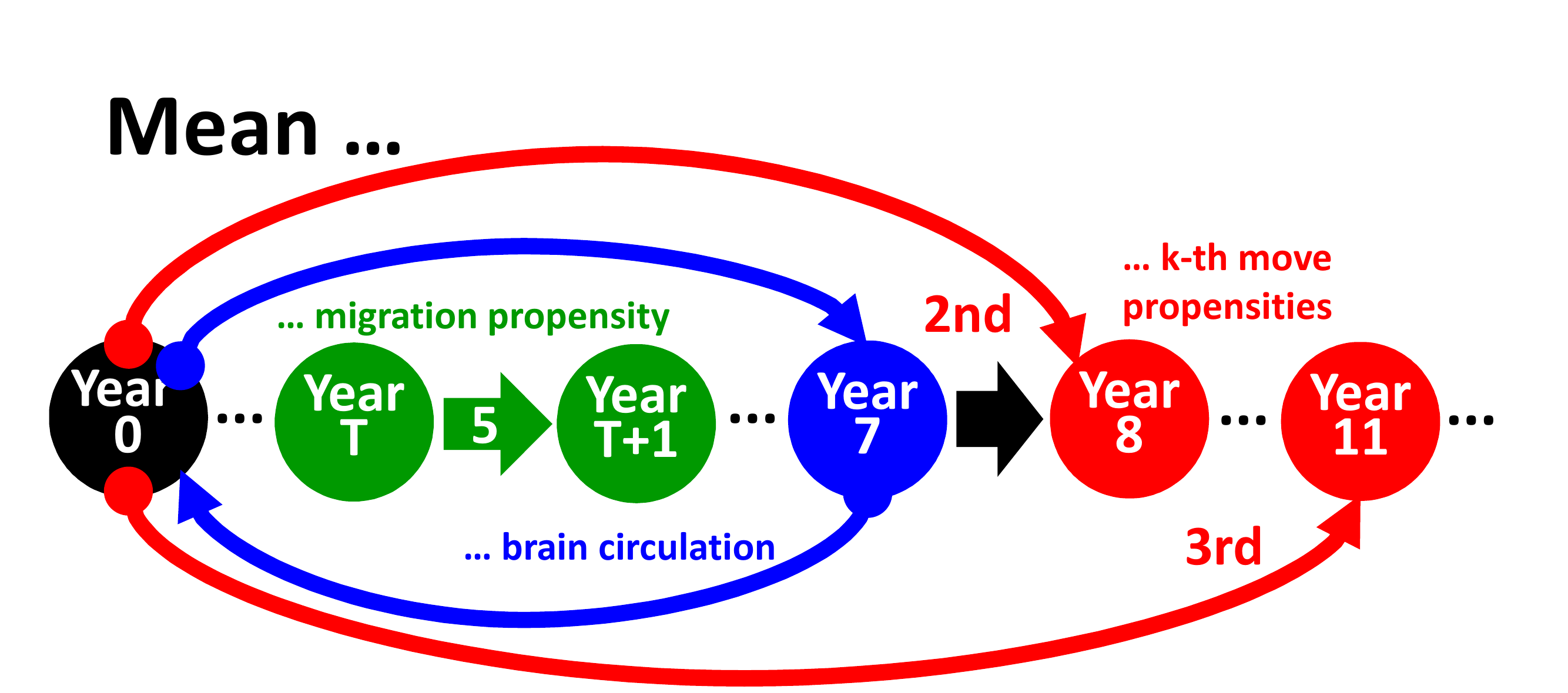}
\end{center}
\caption{Prototypical migration career of a computer scientist according to the WWW. Shown are the mean values for ($k$th move) propensities and brain circulation. That is, on average a scientist makes the next move after 5 years (green). Making two moves takes on average 8 years (red), and three moves 11 years (red). She moves back to her roots, if at all, after 8 years (blue). (Best viewed in color)\label{fig:means}}
\end{figure}

Specifically, we described the creation of GeoDBLP that, in contrast to existing 
migration research, involved propagation of only few seed locations across bibliographic data, namely 
the DBLP network of authors and papers. The result was a database of over 5 million
unique author-paper-pairs mostly labeled with geo-tags, which was used for a detailed statistical analysis. 
The statistical regularities and patterns discovered are encouraging: 
we could estimate statistical regularities for migration propensities that align well but actually go beyond knowledge in the migration and scientometric
 literature --- typically concluded from small-scale, unregistered data 
only --- and establish for the first time that there are no cultural boundaries for the timing 
events underlying migration. The statistical regularity remains similar no matter what country 
you are looking at. Thus, moving on to a new position is a common pattern in terms of timing 
across different countries from the US to China over Germany, and Australia 
and independent of geography, ideology, politics or religion. 
The resulting prototypical migration career is sketched in Fig.~\ref{fig:means}.
This is interesting, since, if nations want to get back their high-level personnel, they have to do that just before the second move, on average in the 7th year.
Otherwise, it is likely that the high-level personnel does not come back anymore.
And recall that only $3\%$ of all scientists actually return. 
If you miss this opportunity, you will have to invest much more, since moving in later stages in a career is memoryless; there is no pressure for high-level personnel to move.
On average scientists move every $5$ years. This high value is due to dominance of researchers in early
academic career stages. For senior scientists, that are the minority, this turns 
into a gamma distribution. For instance, we make two moves within $8$ years on average, while making three moves takes on average $11$ years. 
Analyzing the author-migration graph reveals for instance that 
China is the largest migration hub in the world, whereas Singapore is the largest 
migration authority. Generally, the east cost of the US receives and sends out researchers; 
the east cost is an incubator. In contrast, the west coast of the US is large migration
authority, probably due to strong new economy and better climate. 
People have had this suspicion but we are showing on a very large scale that this insights go beyond folklore.

In general, our findings suggest that the WWW, together with data mining to deal with missing information, may complement existing migration data sources, resolve inconsistencies arising from different definitions of migration, and provide new and rich information on migration patterns of computer scientists. 
However, a lot remains to be done. 
One should monitor migration over time and validate gravity models for international migration. 
One should also investigate the distribution over distances traveled when migrating. It is certainly
more complex and most likely follows a mixtures of distributions. Initial results show that there
are several modes, indicating that there are cultural boundaries. 
Other interesting avenues for future work are geographical topic models to discover research trends 
across the world and to realize expert finding systems that know  where the experts are at any time.
The most promising direction is to extend our results beyond computer science. 

Nevertheless, our results are an encouraging sign that harvesting and inferring data from the web at large-scale may give fresh impetus to demographic research; 
we have only started to look through the world-wide web lens onto it.

{\bf Acknowledgments:} 
This work was partly funded by the Fraunhofer ATTRACT fellowship ``STREAM'' and by the DFG, KE 1686/2-1.

\bibliographystyle{abbrv}
\bibliography{references}

\begin{thebibliography}{10}

\bibitem{aitchison57lognormal}
J.~Aitchison and J.~Brown.
\newblock {\em The Lognormal Distribution}.
\newblock Cambridge University Press, 1957.

\bibitem{barabasi05nature}
A.~Barabasi.
\newblock The origin of bursts and heavy tails in human dynamics.
\newblock {\em Nature}, 435:207--211, 2005.

\bibitem{barabasi12nature}
A.~Barabasi, C.~Song, and D.~Wang.
\newblock Handful of papers dominates citation.
\newblock {\em Nature}, 491:267--270, 2012.

\bibitem{bengio06lp}
Y.~Bengio, O.~Delalleau, and N.~{Le Roux}.
\newblock Label propagation and quadratic criterion.
\newblock In O.~Chapelle, B.~Sch{\"o}lkopf, and A.~Zien, editors, {\em
  Semi-Supervised Learning}, pages 193--216. {MIT} Press, 2006.

\bibitem{bohorquez09nature}
J.~Bohorquez, S.~Gourley, A.~Dixon, M.~Spagat, and N.~Johnson.
\newblock Common ecology quantifies human insurgency.
\newblock {\em Nature}, 462:911--914, 2009.

\bibitem{cohen08pnas}
J.~Cohen, M.~Roig, D.~Reuman, and C.~GoGwilt.
\newblock International migration beyond gravity: A statistical model for use
  in population projections.
\newblock {\em PNAS}, 105(40):15269--15274, Oct. 2008.

\bibitem{editorial10nature}
Editorial.
\newblock Singapore's salad days are over.
\newblock {\em Nature}, 468(731), 2010.

\bibitem{gabaix03nature}
X.~Gabaix, G.~Parameswaran, V.~Plerou, and H.~Stanley.
\newblock A theory of power law distributions in financial market fluctuations.
\newblock {\em Nature}, 423:267--270, 2003.

\bibitem{wissenschaftsrat10}
{German Council of Science and Humanities}.
\newblock Recommendations on german science policy in the european research
  area.
\newblock Technical Report Drs. 9866-10, Berlin 02 07 2010, 2010.

\bibitem{goetz09icwsm}
M.~Goetz, J.~Leskovec, M.~McGlohon, and C.~Faloutsos.
\newblock Modeling blog dynamics.
\newblock In {\em Proc. of the Third International Conference on Weblogs and
  Social Media (ICWSM)}, 2009.

\bibitem{gonzales08nature}
M.~Gonzales, C.~Hidalgo, and A.~Barabasi.
\newblock Understanding individual human mobility patterns.
\newblock {\em Nature}, 453:779--782, 2008.

\bibitem{hidalgo08physica}
C.~Hidalgo and C.~{Rodriguez-Sickert}.
\newblock The dynamics of a mobile phone network.
\newblock {\em Physica A}, 287(12):3017--3024, 2008.

\bibitem{hui10hot}
P.~Hui, R.~Mortier, M.~Piorkowski, T.~Henderson, and J.~Crowcoft.
\newblock Planet-scale human mobility measurement.
\newblock In {\em Proc. of the 2nd ACM International Workshop on Hot Topics on
  Planet-Scale Measurements}, 2010.

\bibitem{kleinberg99}
J.~Kleinberg.
\newblock Authoritative sources in a hyperlinked environment.
\newblock {\em J. ACM}, 46(5):604--632, Sept. 1999.

\bibitem{lam10}
W.~Lam.
\newblock China's brain drain dilemma: Elite emigration.
\newblock {\em The Jamestown Foundation: China Brief Volume}, 10(16), 2010.

\bibitem{lee11science}
R.~Lee.
\newblock The outlook for population growth.
\newblock {\em Science}, 333(6042):569--573, 2011.

\bibitem{leskovec09kdd}
J.~Leskovec, L.~Backstrom, and J.~Kleinberg.
\newblock Meme-tracking and the dynamics of the news cycle.
\newblock In {\em Proc. of the 15th ACM SIGKDD International Conference on
  Knowledge Discovery and Data Mining (KDD)}, pages 497--506, 2009.

\bibitem{leskovec08www}
J.~Leskovec and E.~Horvitz.
\newblock Planetary-scale views on a large instant-messaging network.
\newblock In {\em Proc. of the 17th International Conference on World Wide Web
  (WWW)}, pages 915--924, 2008.

\bibitem{ley09}
M.~Ley.
\newblock {DBLP}: some lessons learned.
\newblock {\em Proc. VLDB Endow.}, 2(2):1493--1500, Aug. 2009.

\bibitem{limpert01bioscience}
E.~Limpert, W.~Stahel, and M.~Abbt.
\newblock Log-normal distributions across the sciences: Keys and clues.
\newblock {\em BioScience}, 51(5):341--352, 2001.

\bibitem{mantegna95nature}
R.~Mantegna and H.~Stanley.
\newblock Scaling behaviour in the dynamics of an economic index.
\newblock {\em Nature}, 376:46--49, 1995.

\bibitem{noulas11wscm}
A.~Noulas, S.~Scellato, C.~Mascolo, and M.~Pontil.
\newblock An empirical study of geographic user activity patterns in
  foursquare.
\newblock In {\em Proc. of the the Fifth International AAAI Conference on
  Weblogs and Social Media (WSCM)}, pages 570--573, 2011.

\bibitem{Page99}
L.~Page, S.~Brin, R.~Motwani, and T.~Winograd.
\newblock The pagerank citation ranking: Bringing order to the web.
\newblock Technical Report 1999-66, Stanford InfoLab, November 1999.

\bibitem{rees09}
P.~Rees, J.~Stillwell, P.~Boden, and A.~Dennett.
\newblock A review of migration statistics literature.
\newblock In {\em UK Statistics Authority, Migration Statistics: the Way Ahead,
  Report 4, Part 2}. UK Statistics Authority, London. ISBN: 978-1-85774-904-5,
  2009.

\bibitem{nsf07}
M.~Regets.
\newblock Research issues in the international migration of highly skilled
  workers: A perspective with data from the united states.
\newblock Technical Report Working Paper SRS 07-203, Arlington, VA: Division of
  Science Resources Statistics, National Science Foundation, 2010.

\bibitem{gibrat30}
R.Gibrat.
\newblock Une loi des repartitions economiques: L'effet proportionelle.
\newblock {\em Bulletin de Statistique General}, 19:469--514, 1930.

\bibitem{Siminil12nature}
F.~Simini, M.~Gonzalez, A.~Maritan, and A.~Barabasi.
\newblock A universal model for mobility and migration patterns.
\newblock {\em Nature}, 484:96--100, 2012.

\bibitem{stewart94info}
J.~Stewart.
\newblock The poisson-lognormal model for bibliometriy/scientometric
  distribitions.
\newblock {\em Information Processing and Management}, 30(2):239--251, 1994.

\bibitem{stillwell05}
J.~Stillwell.
\newblock Inter-regional migration modelling: A review and assessment.
\newblock In {\em Proc. of the 45th Congress of the European Regional Science
  Association}, 2005.

\bibitem{un2012}
{UN System Task Team on the Post-2015 UN Development Agenda}.
\newblock Migration and human mobility.
\newblock Technical report, 2012.

\bibitem{wang11kdd}
D.~Wang, D.~Pedreschi, C.~Song, F.~Giannotti, and A.~Barabasi.
\newblock Human mobility, social ties, and link prediction.
\newblock In {\em Proceedings of the ACM SIGKDD International Conference on
  Knowledge Discovery and Data Mining (KDD)}, 2011.

\bibitem{zagheni12websci}
E.~Zagheni and I.~Weber.
\newblock You are where you e-mail: using e-mail data to estimate international
  migration rates.
\newblock In {\em Proc. of the ACM Conference on Web Science (WebSci)}, pages
  348--351, 2012.

\bibitem{zhu02}
X.~Zhu and Z.~Ghahramani.
\newblock Learning from labeled and unlabeled data with label propagation.
\newblock Technical Report CMU-CALD-02-107, Carnegie Mellon University, 2002.

\bibitem{zipf46asr}
G.~Zipf.
\newblock The $p_1p_2\slash d$ hypothesis: On the inter-city movement of
  persons.
\newblock {\em Am. Sociol. Rev.}, 11:677---686, 1946.

\end{thebibliography}

\end{document}